\documentclass[twocolumn,trackchanges,onecolappendix]{aastex631}

\usepackage{amsmath}
\usepackage{makecell}
\usepackage{graphicx}
\usepackage{amsthm,amsmath,amssymb} 
\usepackage{mathrsfs}
\usepackage{longtable}
\usepackage{subfigure}
\usepackage{rotating}
\usepackage{amssymb}
\usepackage{epsfig}
\usepackage{footmisc}
\usepackage{multirow}
\usepackage{enumerate}
\usepackage{enumitem}
\usepackage{float}
\usepackage[section]{placeins}
\usepackage{url}
\usepackage{soul}
\usepackage{CJK}
\usepackage{booktabs}
\usepackage{comment}
\usepackage{xcolor}
\usepackage[normalem]{ulem}

\begin{document}


\title{Indication for Decreasing Dispersion Measure in the Population of Repeating Fast Radio Bursts \\ and  Connection to Young Supernova Remnant Expansion}

\author[0000-0002-6165-0977]{Xiang-han Cui}
\affiliation{National Astronomical Observatories, Chinese Academy of Sciences, Beijing 100101, China}
\affiliation{ASTRON, the Netherlands Institute for Radio Astronomy, Oude Hoogeveensedijk 4,7991 PD Dwingeloo, The Netherlands}
\email{cuixianghan@nao.cas.cn}

\author[0000-0003-1908-2520]{Cheng-min Zhang}
\affiliation{National Astronomical Observatories, Chinese Academy of Sciences, Beijing 100101, China}
\affiliation{School of Astronomy and Space Science, University of Chinese Academy of Sciences, Beijing 100049, China}
\affiliation{School of Physical Sciences, University of Chinese Academy of Sciences, Beijing 100049, China}

\author[0000-0003-3010-7661]{Di Li}
\affiliation{New Cornerstone Science Laboratory, Department of Astronomy, Tsinghua University, Beijing, 100084, China}
\affiliation{National Astronomical Observatories, Chinese Academy of Sciences, Beijing 100101, China}

\author[0000-0002-3386-7159]{Pei Wang}
\affiliation{National Astronomical Observatories, Chinese Academy of Sciences, Beijing 100101, China}
\affiliation{School of Astronomy and Space Science, University of Chinese Academy of Sciences, Beijing 100049, China}

\author{Erbil G\"{u}gercino\u{g}lu}
\affiliation{School of Arts and Sciences, Qingdao Binhai University, Huangdao District, Qingdao, 266555, China.}
\affiliation{National Astronomical Observatories, Chinese Academy of Sciences, Beijing 100101, China}

\author[0000-0001-8503-6958]{Joeri van Leeuwen}
\affiliation{ASTRON, the Netherlands Institute for Radio Astronomy, Oude Hoogeveensedijk 4,7991 PD Dwingeloo, The Netherlands}

\author[0000-0002-7372-4160]{Yi-dan Wang}
\affiliation{National Astronomical Observatories, Chinese Academy of Sciences, Beijing 100101, China}
\affiliation{School of Astronomy and Space Science, University of Chinese Academy of Sciences, Beijing 100049, China}

\author[0000-0002-9390-9672]{Chao-wei Tsai}
\affiliation{National Astronomical Observatories, Chinese Academy of Sciences, Beijing 100101, China}
\affiliation{Institute for Frontiers in Astronomy and Astrophysics, Beijing Normal University,  Beijing 102206, China}
\affiliation{School of Astronomy and Space Science, University of Chinese Academy of Sciences, Beijing 100049, China}

\author[0000-0001-5738-9625]{Xiang-lei Chen}
\affiliation{National Astronomical Observatories, Chinese Academy of Sciences, Beijing 100101, China}

\author[0009-0002-7690-7381]{Wen-qi Ma}
\affiliation{Xinjiang Astronomical Observatory, Chinese Academy of Sciences, Urumqi 830011, China}
\affiliation{School of Astronomy and Space Science, University of Chinese Academy of Sciences, Beijing 100049, China}

\author[0000-0002-5815-6548]{Habtamu Menberu Tedila}
\affiliation{National Astronomical Observatories, Chinese Academy of Sciences, Beijing 100101, China}

\begin{abstract}
Fast Radio Bursts (FRBs) are millisecond-duration, highly energetic radio transients of uncertain origin.
Repeating FRBs provide an excellent population for investigating their nature, particularly through studies of parameter evolution.
Out of the 63 repeaters monitored by CHIME, 
we select the 19 sources with more than 10 detected bursts, 
and examine their long-term dispersion measure (DM) evolution. 
Seven sources show statistically significant DM evolution and are classified as the golden sample. 
Of these, five exhibit a decreasing DM trend and two show an increasing trend.
We then perform a binomial test under the null hypothesis that decreasing and increasing DM variation trends have equal probabilities.
The current combined sample, including our golden sample and additional repeaters with reported DM change rate from the literature, gives a p-value of 0.033, supporting that decreasing DM trends are more common in the repeating FRB population.
This statistical result is consistent with scenarios in that the local electron density around repeaters generally decreases with time, for example, due to expansion of a young supernova remnant (SNR).
Finally, within the SNR expansion model, we provide an illustrative estimate of the SNR contributions to the DM for different ejecta masses. 

\end{abstract}

\keywords{Radio transient sources (2008); Radio bursts (1339); Supernova remnants (1667); Astrostatistics (1882)}

\section{Introduction} \label{1}
Our understanding of millisecond-duration radio flashes at cosmological distances, known as fast radio bursts (FRBs), have made significant progress in both observations and theory since their first systematic study in 2007  \citep{Lorimer07, Cordes19, Petroff22, ZhangB23}.
Although FRB-like signals have been detected from a Galactic magnetar SGR 1935+2154 \citep{CHIME20a, Bochenek20} and some repeating FRBs are associated with persistent radio sources (PRSs) \citep{Chatterjee17, NiuCH22}, their exact origin remains unclear. 
Continuous monitoring of FRBs is crucial for understanding their origin and emission mechanisms \citep{Bailes22}.
The Five-hundred-meter Aperture Spherical radio Telescope (FAST; \citealt{NanRD11, LiD18}), 
with its exceptional sensitivity, has conducted extensive observations of several high burst rate repeating FRBs and accumulated a vast amount of data \citep{LiD21}. 
The Canadian Hydrogen Intensity Mapping Experiment (CHIME), thanks to its wide field of view \citep{CHIME18}, has discovered a large number of FRB sources and continues to monitor them \citep{CHIME21a}.
These rich datasets provide strong support for statistical studies of FRB population properties and their physical characteristics, such as population diversity \citep{Fonseca20, Gardenier2021, CuiXH21}, energy or luminosity constraints \citep{LuoR18, LiD21, CuiXH22, Wang2024}, spectral statistics \citep{Macquart18, CuiXH25}, and cosmological applications \citep{Macquart20, James22, Connor25}. 

Continuous monitoring of repeaters is important for uncovering their physical properties.
Regular observations of FRB 20180916B \citep{CHIME20b} and FRB 20121102A \citep{Rajwade20} have revealed activity periods of approximately 16 and 157 days, respectively, suggesting that their activity may be modulated by source surroundings \citep{LanHT24}.
Several high-rate repeaters, such as FRB 20121102A \citep{Oostrum2020,LiD21}, FRB 20190520B \citep{NiuCH22}, FRB 20201124A \citep{ZhangYK22}, FRB 20220912A \citep{ZhangYK23}, and FRB 20240114A \citep{ZhangJS25}, show bimodal or complex energy distributions with high completeness, indicating that multiple mechanisms may be involved in triggering or powering active repeaters.
Based on more than 10,000 bursts from FRB 20240114A, \cite{ZhangJS25} placed constraints on the energy budget for the magnetar-powered model.
With precise localization, four repeating FRBs have been found to be associated with persistent radio sources (PRSs): FRB 20121102A \citep{Chatterjee17}, FRB 20190520B \citep{NiuCH22}, FRB 20240114A \citep{Bruni25}, and FRB 20190417A \citep{Moroianu25}.
In addition, FRB 20201124A has been reported to be associated with a PRS candidate \citep{Ravi22, Bruni24}.
PRSs may be related to remnants of massive star core collapse, which has important implications for constraining their progenitors.
Moreover, statistical analyses on rotation measures (RMs) have revealed frequency-dependent RM scattering \citep{FengY22} and RM variations \citep{Michilli18, XuH22, LiY2026}, suggesting that active repeaters may experience multi-path scattering during propagation related to highly complex surrounding environments \citep{ZhangJL2026}.
Frequency-dependent depolarization was first noted in repeating sources and has been reported for some apparently non-repeating sources (hereafter non-repeaters) \citep{Uttarkar26}, which may be related to propagation effects in complex environments.
It remains unclear whether these non-repeaters are undetected repeaters or belong to a different population that shares similar environments with repeaters.

Taken together, these observational results provide valuable clues to the environments and progenitors of FRBs.
Since the dispersion measure (DM) is an integral of the electron density along the line of sight \citep{Lorimer12}, it provides a powerful probe of the path. 
In particular, DM variations observed on yearly timescales are unlikely to originate from large-scale structures such as the Milky Way ($\rm DM_{MW}$), the intergalactic medium ($\rm DM_{IGM}$), or host galaxies ($\rm DM_{host}$), instead contributions from the source surroundings \citep{YangYP17}.
Galactic radio pulsars show evidence that their local environment affects the observed DM.
This happens in two ways: by an additional  dispersion measure due to the local material and by temporal variations, which are thought to be inhomogeneities and motions in this material. 
First, \citet{Straal20} showed that pulsars in pulsar wind nebulae (PWNe) and supernova remnants (SNRs)
have higher DM than pulsars unassociated with such remnants.
Next, in several pulsars, DM variations have been observed on timescales of years \citep[e.g.][]{Kuzmin08, Petroff13, McKee18}, likely associated with expanding or moving PNWe, SNRs, or filaments.

The DM variations in FRBs are likely to be caused by the local environments too, which offers an excellent opportunity to probe their surroundings.
For FRB 20180916B, a linear fit to the measured DMs yield a slope of $\sim 0.03\ \pm 0.12\, \rm pc\ cm^{-3}\ yr^{-1}$ at 1$\sigma$ error confidence level \citep{WangYB23}, indicating that its DM trend is not statistically significant.
Long-term monitoring of several repeating FRBs has uncovered diverse DM evolution behavior (quoted errors denote 1$\sigma$):
\begin{itemize}[leftmargin=*, noitemsep, topsep=0pt, parsep=0pt, partopsep=0pt]
\item FRB 20180301A: $-2.7 \pm 0.2\, \rm pc\, cm^{-3}\, yr^{-1}$  \citep{Kumar23};
\item FRB 20121102A: $-3.93 \pm 0.11\, \rm pc\, cm^{-3}\, yr^{-1}$ \citep{WangP25}, or decreasing $\sim 25\, \rm pc\, cm^{-3}$ in the past five years \citep{Snelders25};
\item FRB 20190520B: $-12.4 \pm 0.3\, \rm pc\, cm^{-3}\, yr^{-1}$  \citep{NiuCH26};
\item FRB 20220529A: $-0.881 \pm 0.001\, \rm pc\, cm^{-3}\, yr^{-1}$  \citep{Pandhi26}\footnote{Preprint posted while the current paper was under review.}.
\end{itemize}
\citet{Curtin25} analyzed 35 repeating FRBs detected by CHIME and found no statistically significant evidence for a consistent long-term trend in DM over timescales of $\sim$2–4 yr.

Most previous studies have focused on individual repeaters or small samples.
The growing number of repeating FRBs now enables a further statistical study of DM variations and their population properties.
In this work, we analyze the DM variations of repeaters reported by CHIME and discuss their physical interpretations.
The manuscript is organized as follows.
In Section \ref{2}, we provide an introduction to CHIME repeaters and describe our data selection.
In Section \ref{3}, we present the fitting methods applied to DM excess and show the statistical results.
In Section \ref{4}, we discuss possible explanations for overall trends in DM.
Finally, a conclusion is presented in Section \ref{5}.

\section{Data Preparation} \label{2}

Around 100 repeating FRBs have been published to date\footnote{https://blinkverse.zero2x.org}, most of which were discovered by CHIME with its wide field of view \citep{CHIME26}.
To ensure the completeness of the dataset, this work focuses on the 63 repeaters detected by CHIME (1252 bursts, up to February 2026)\footnote{https://www.chime-frb.ca/repeaters}.
Each recorded event includes a burst ID, UTC time, sky position (RA and Dec), DM, DM error, and signal-to-noise ratio (S/N). 
A small fraction additionally provides fluence, flux, width, scattering time, and dynamic spectrum.
Recently, CHIME/FRB Catalog 2 (hereafter Catalog 2) has been released \citep{CHIME26}, however we do not use it here, because it contains data only up to 2023 and includes fewer repeaters' bursts (981 in total). 

Among the 63 repeaters, we focus on 19 sources with more than ten detected bursts for subsequent analysis of DM variations. 
Then, we check the S/N of total 1252 bursts (19 repeaters) and confirmed that all of them are larger than 7, thus, ensuring their suitability for analysis.
Among these 19 active repeaters, FRB 20220912A exhibits the highest activity, with 445 detected bursts, while FRB 20201130A displays the lowest activity, with 12 bursts. 
The remaining sources lie between these two, with a median burst count of 22.
 Detailed information on the selected sources is listed in Appendix \ref{A}.

DM represents the integrated effect of the electron density ($n_e$) along the propagation path ($d$) \citep{Lorimer12}
\begin{equation}
    {\rm DM} \equiv \int^{d}_{0}n_e {\rm d}l.
\label{dm-def}
\end{equation}
Considering the differences in electron density among various components along the path, the DM of one FRB consists of at least four parts \citep{Cordes19}:
\begin{equation}
    {\rm DM = DM_{MW} + DM_{halo} + DM_{IGM}} + \frac{\rm DM_{host}}{1+z},
\label{dm-all}
\end{equation}
where ${\rm DM_{MW}}$, ${\rm DM_{halo}}$, ${\rm DM_{IGM}}$, and ${\rm DM_{host}}$ represent DM contributions from the Milky Way galaxy, Galactic halo, intergalactic medium (IGM), and host galaxy, respectively.
${\rm DM_{host}}$ can also be separated into contributions from its interstellar medium (ISM) (${\rm DM_{host,ISM}}$) and the local environment of the source (${\rm DM_{local}}$).
If we assume that, on timescales of a few years, large-scale structures such as the Galactic halo, the IGM, and the ISM of the host galaxy remain nearly constant, therefore the observed variations in DM most likely reflect changes in the local environment \citep{YangYP17, WangP25}.
The Milky Way contribution ${\rm DM_{MW}}$ can be estimated using Galactic electron density models such as YMW16 \citep{YaoJM16}, which account for more  Galactic structures.
We define ${\rm DM_{exc}}$ to represent the components that exceed the Galactic contribution
\begin{equation}
    {\rm DM_{exc} = DM_{halo} + DM_{IGM}} + \frac{\rm DM_{host,ISM}+DM_{local}}{1+z}.
\label{dm-exc}
\end{equation}
In the following statistical analysis, we examine ${\rm DM_{exc}}$ for each repeater. 

\section{Statistical Results} \label{3}
In this section, we present the calculation method in detail and show DM variation for each repeater as well as their uncertainties.
\subsection{Calculation method} \label{3.1}
Since the arrival time of each burst is known for all sources, we plot the distribution of the measured ${\rm DM_{exc}}$ as a function of time, including their errors.
When fitting the overall trend, we apply a linear rather than a higher-order polynomial fit.
There are two reasons for this choice: (1) the burst numbers for several repeaters are limited, which is insufficient to robustly constrain more complex variations such as oscillatory features; 
(2) part of the apparent variability in FRBs may be influenced by instrumental effects such that higher-order fits could introduce misleading terms.
Therefore, we employ a linear fit to characterize the overall trend of DM variation.
We first estimate the long-term ${\rm DM_{exc}}$ variation rate using a weighted least-squares (WLS) linear fit.
A Bayesian method is then used to test the inferred DM rate.
For sources where both methods exclude zero at 95\% confidence level and agree on the sign of the trend, we conclude that the evolution is statistically significant.
Only this set of sources (hereafter: the ``golden sample") are used for further analysis. 

Specifically, we adopt a weighted linear fitting approach to estimate the long-term DM variation rate for each repeater. 
From the measurements in pulsars \citep[e.g,][]{Backer_2000} and FRBs \citep[e.g,][]{Pandhi26}, we know that DMs can also display significant stochastic change. 
Our goal—no matter how intriguing they may be—is not to engage in fitting to those, but to focus on long-term evolution. 
To account for the different measurement uncertainties among individual bursts, the fit is performed using weights determined by the measurement uncertainties of individual bursts, such that higher weight is given to bursts with smaller DM errors.
An error inflation factor based on the reduced $\chi^2$ of an initial fit is subsequently applied in the fitting, in order 
to mitigate the potential underestimation of the DM errors. 
Such uncertainty scaling is commonly used in pulsar timing \citep[the EFAC factor; see, e.g.,][]{Iraci2024}. 
All input DM measured errors for each source are multiplied with this EFAC factor, 
${\rm error}_{\rm inflated} = {\rm EFAC} \times {\rm error}_{\rm original}$,
until the reduced $\chi^2$ becomes unity.
We note that all plots continue to display the original CHIME-reported error bars.

Bayesian methods are generally effective for statistical analyses with large datasets.
In this work, the Bayesian approach is therefore only used as a robustness test to validate the results obtained from the WLS fitting.
We adopt a linear model for the DM evolution, simultaneously fitting the slope (DM variation rate) with its intercept and an EFAC inflation factor.
The posterior probability distribution is given by
\begin{equation}
    P(\mathrm{model}\,|\,{\rm DM}_i)
    \propto
    P({\rm DM}_i\,|\,\mathrm{model})
    \times
    P(\mathrm{model}),
\label{Bayesian}
\end{equation}
where $P(\rm{model})$ represents uniform priors on all parameters.
In particular, the prior range for the DM variation rate is from $[-50,50]\ {\rm pc\ cm^{-3}\ yr^{-1}}$, the intercept sets to $[0,3000]\ {\rm pc\ cm^{-3}\ yr^{-1}}$, and the EFAC parameter is allowed within [1,100].
$P({\rm DM}_i\,|\,\mathrm{model})$ represents a Gaussian likelihood.
The posterior is sampled using a Markov Chain Monte Carlo (MCMC) method implemented with the \texttt{emcee} package, and the median values and 95\% confidence level of the DM rate are derived.
The MCMC sampling is performed 20 walkers with 5000 steps for parameter estimation.
We note that this Bayesian approach is based on the same underlying linear model as the WLS method, and the priors of several free parameters are  uniform. 
Therefore, the Bayesian results are expected to fluctuate around those obtained with the WLS method and do not constitute a fully independent cross-check of WLS.

\begin{figure}[ht]
\plotone{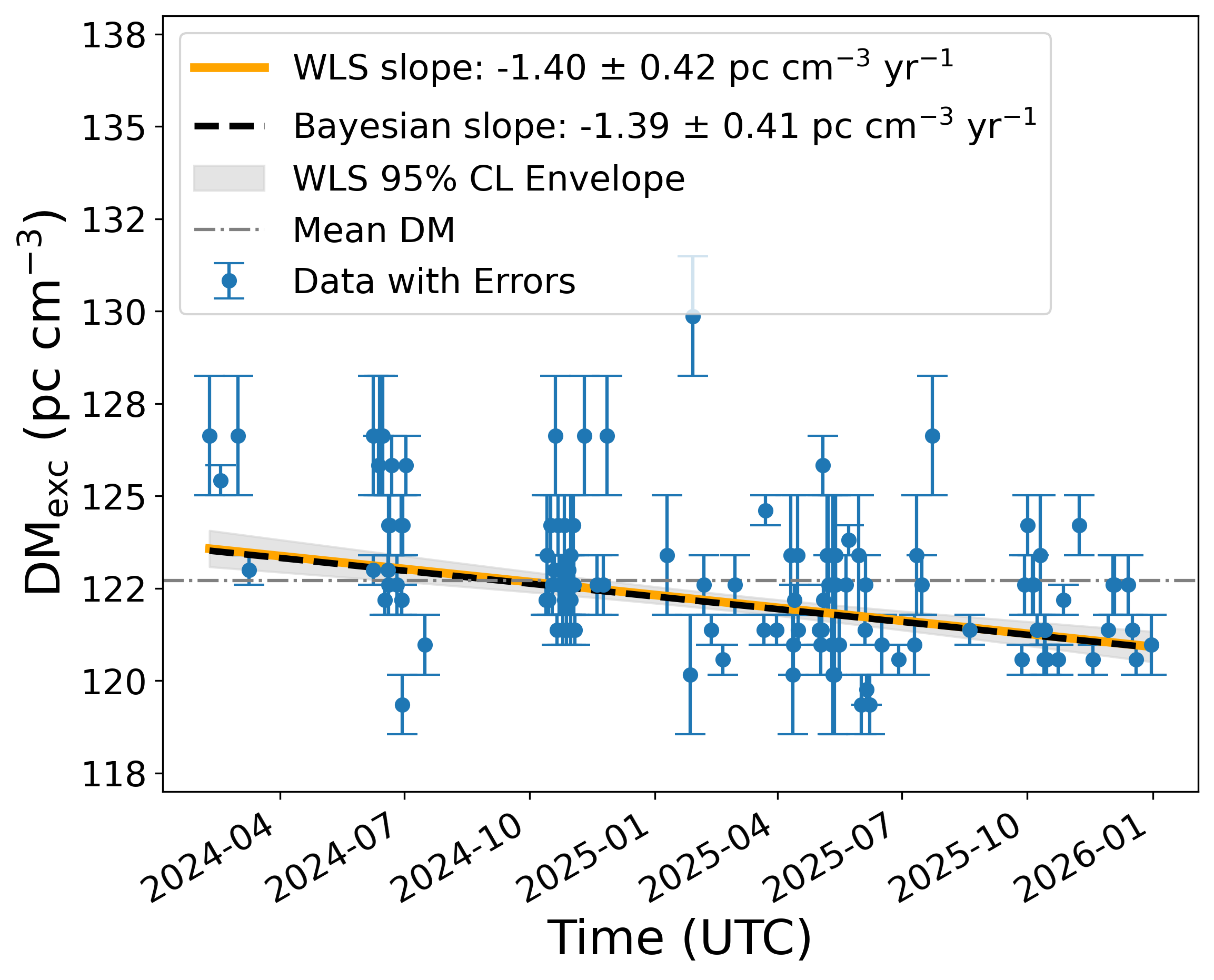}
\caption{A clear case where a DM rate of change can be established (FRB 20240209A).
Blue points represent individual $\rm DM_{exc}$ with original errors. 
The orange solid line and black dashed line represent the best-fits from weighted least-squares (WLS) and Bayesian method, respectively, with their 95\% confidence level (CL).
The gray shaded region shows the 95\% CL envelope from the WLS fit.
The gray dashed-dot line indicates the mean value of $\rm DM_{exc}$ of the source.}
\label{dm_trend_example}
\end{figure}

We subsequently define a subset of sources as the golden sample, as described above. 
For the remaining sources, DM evolution cannot be completely ruled out, due to the limited number of bursts.
But to avoid over-interpretation, we do not further analyse these non-golden-sample sources.
In Figure \ref{dm_trend_example} we show  an example of our procedure for estimating the DM variation rate, for the case FRB 20240209A.

\subsection{$\rm DM$ variation trends of repeaters}

\begin{table*}
\centering
\caption{Properties and DM variation rates of the golden-sample repeating FRBs.}
\label{golden_sample}
\begin{tabular}{lccccccccc}
\hline
\hline
FRB & Burst & RA & Dec & $l$ & $b$ & Mean $\rm DM_{exc}$ &
WLS rate & Bayesian rate \\
Name &Number & (deg) & (deg) & (deg) & (deg) & (pc cm$^{-3}$) &
(pc cm$^{-3}$ yr$^{-1}$) & (pc cm$^{-3}$ yr$^{-1}$)\\
\hline
\multicolumn{8}{l}{\textit{Sources with decreasing DM}} \\
FRB\,20240209A & 127 & 289.89 & 86.07 & 118.57 &  26.58 & 122.71 &
$-1.40\pm0.42$ & $-1.39\pm 0.41$ \\
FRB\,20190303A &  38 & 208.25 & 48.25 &  97.48 &  65.72 & 200.99 &
$-0.23\pm 0.13$ & $-0.22\pm 0.12$ \\
FRB\,20240316A$^{a}$ &  34 & 354.58 & 32.38 & 105.47 & $-$28.00 & 308.61 &
$-14.78\pm 8.82$ & $-15.16^{+8.18}_{-8.51}$ \\
FRB\,20210323C &  17 & 122.00 & 72.33 & 142.59 &  31.54 & 239.65 &
$-1.28\pm 0.30$ & $-1.28^{+0.33}_{-0.32}$ \\
FRB\,20181128A &  16 &  74.00 & 63.38 & 146.62 &  12.43 & 297.36 &
$-0.89\pm 0.20$ & $-0.89^{+0.20}_{-0.21}$ \\
\hline
\multicolumn{8}{l}{\textit{Sources with increasing DM}} \\
FRB\,20220912A & 445 & 209.05 & 48.70 &  157.05 & 61.64 & 201.4 &
$1.41\pm 0.53$ & $1.41\pm 0.53$ \\
FRB\,20190117A & 27 & 331.75 & 17.38 &  76.35 & $-$30.25 & 356.97 &
$0.80\pm 0.19$ & $0.80\pm 0.19$ \\
\hline
\hline
\end{tabular}
\begin{flushleft}
Notes. Bursts are taken from the CHIME/FRB repeater catalog
(https://www.chime-frb.ca/repeaters).
DM variation rates are measured using weighted least-squares (WLS) and Bayesian methods, and uncertainties quoted for both methods correspond to the 95\% confidence level.
$^{a}$ The uncertainty of the fitted rate for this source is likely to be underestimated due to the high leverage of an isolated point within the limitations of the EFAC-only approach.
\end{flushleft}
\end{table*}

\begin{figure*}
\plotone{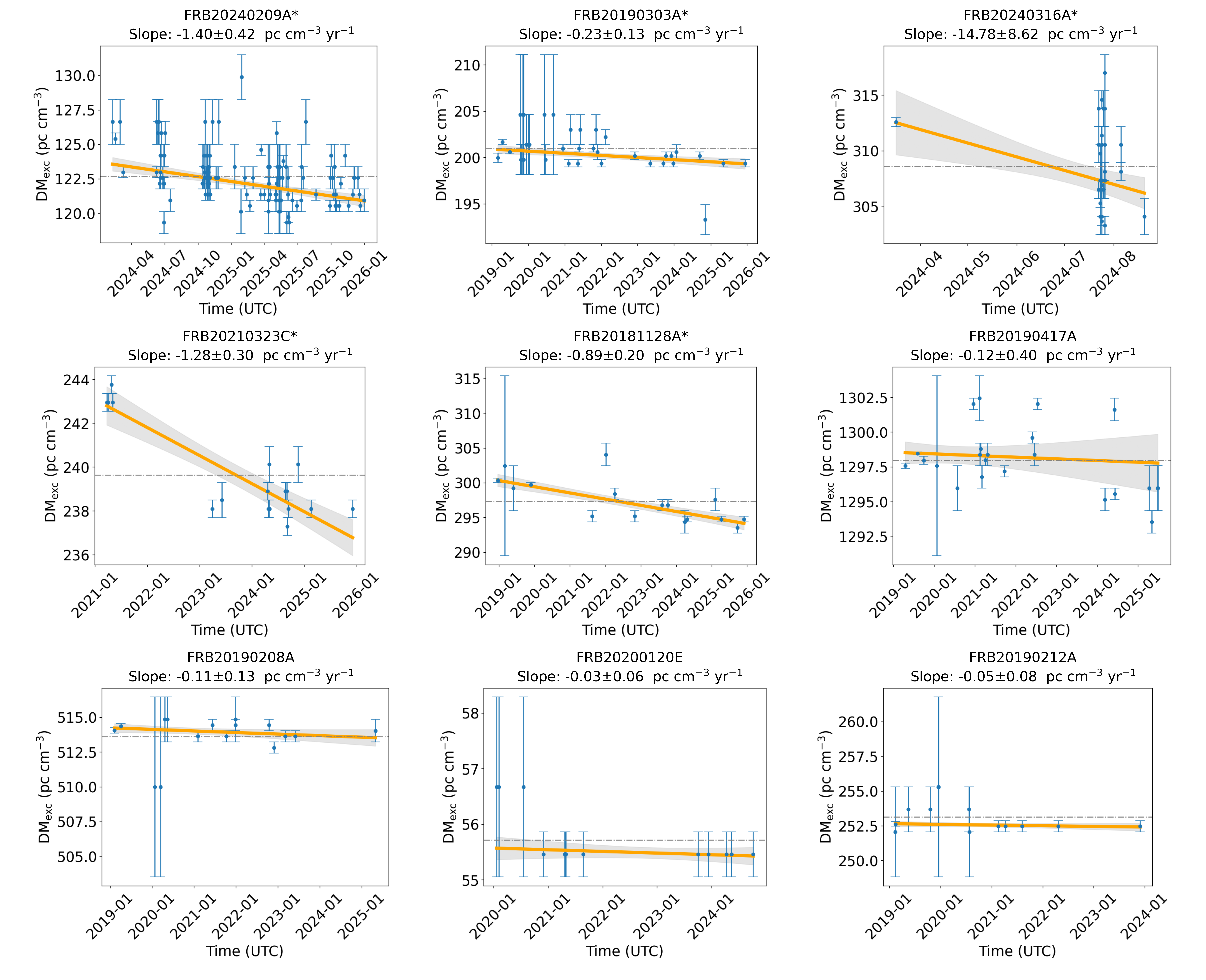}
\caption{A group of all repeaters with an overall decreasing $\rm DM_{exc}$ trend.
Sources marked with an asterisk ``*” are considered golden samples.
Other annotations are the same as in Figure \ref{dm_trend_example}.}
\label{frb_slope_grid_neg}
\end{figure*}

\begin{figure*}
\plotone{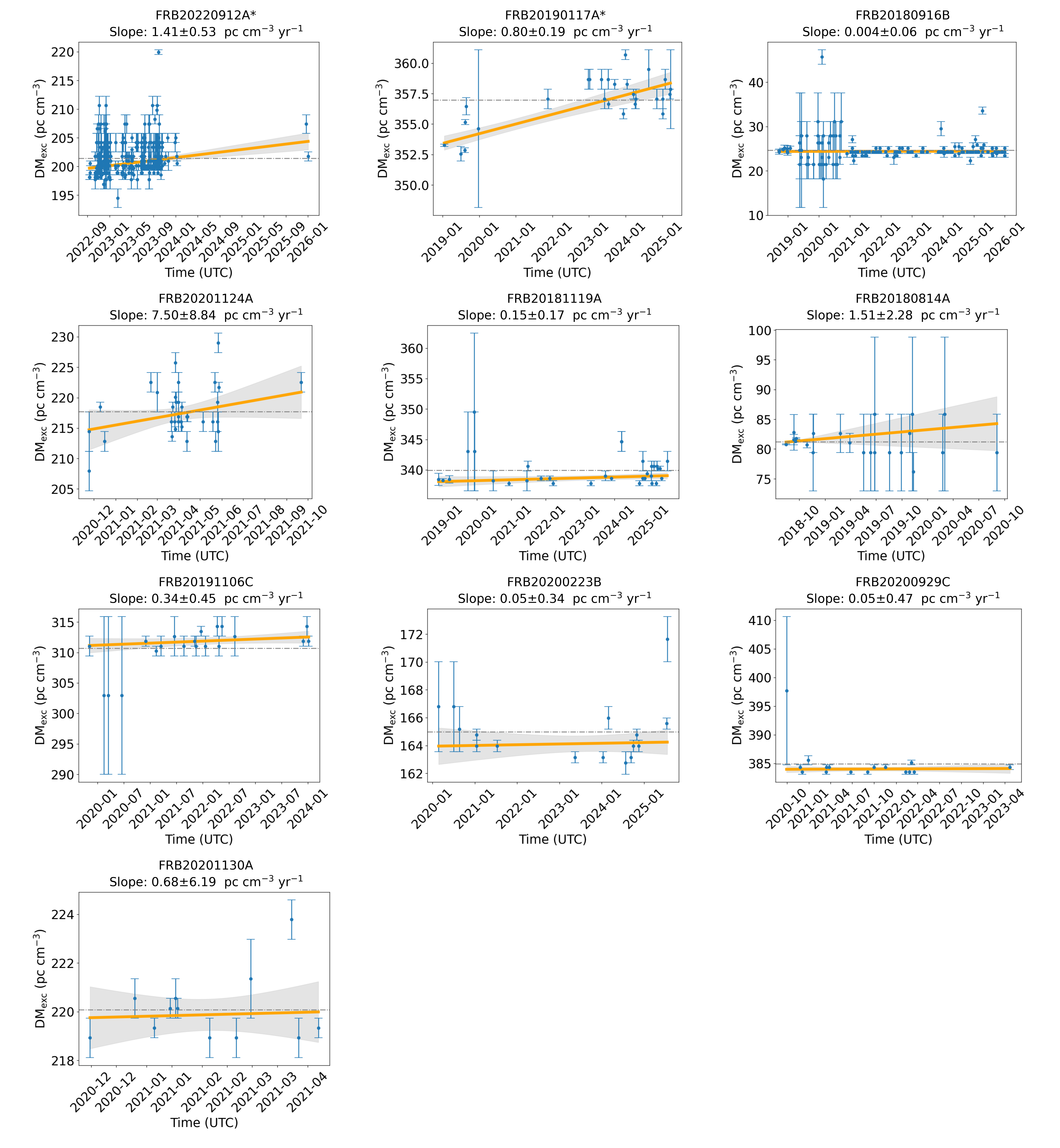}
\caption{A group of all repeaters with an overall increasing $\rm DM_{exc}$ trend.
Sources marked with an ``*” belong golden samples.
Other annotations are the same as in Figure \ref{dm_trend_example}.}
\label{frb_slope_grid_pos}
\end{figure*}

We apply the method in Section \ref{3.1} to the 19 selected repeaters and find that their overall DM variations exhibit two apparent behaviors: a decreasing trend and an increasing trend.
Based on the criteria described above, seven sources satisfy the requirements of the golden sample.
The group figures for sources exhibiting decreasing and increasing DM trends are shown in Figures \ref{frb_slope_grid_neg} and \ref{frb_slope_grid_pos}, respectively, with the golden sample sources marked with an asterisk (*).

In Figure \ref{rate_hist}, we present histograms of the DM variation rates for the golden sample, separated into sources with decreasing and increasing DM trends.
Among the seven golden sample sources, five exhibit decreasing DM trend, and two shows an increasing trend, as listed in Table \ref{golden_sample}.
The median DM rate for the decreasing group is $-1.28{\rm\,pc\ cm^{-3}\ yr^{-1}}$, while that for the increasing group is $1.10{\rm\,pc\ cm^{-3}\ yr^{-1}}$.
To further account for the uncertainties of individual DM rates, we also construct smoothed histograms.
In this approach, the DM rate of each source is represented by a Gaussian distribution centered on its best-fit value, with a width determined by its uncertainty and normalized by its area.
The summed distribution forms a smoothed histogram, which can be interpreted as the probability distribution of the DM variation rates according to current samples.

Below, we compare our results with DM time-series analyses for repeating FRBs  
that overlap with our samples and where previously published\footnote{While our manuscript was under review, a preprint appeared on arXiv on the repeater population in CHIME Catalog 2  \citep{Cook26}.
It reports six repeaters with significant DM variations, four of which show trends consistent with our results (FRB 20220912A, FRB 20190117A, FRB 20210323C, and FRB 20181128A). 
The other two either are not included in our repeater dataset (FRB 20220529A) or feature fewer then 10 bursts and hence do not meet our selection criteria (FRB 20190907A). 
Conversely, two repeaters featured in our work (FRB 20240209A and FRB 20240316A) are not included in \citet{Cook26} because the Catalog 2 data runs up only to September 2023.}.
\cite{WangYB23} analyzed approximately three years of FRB 20180916B data and reported a DM variation rate {$\sim 0.03\ \pm 0.12\, \rm pc\ cm^{-3}\ yr^{-1}$ (1$\sigma$), which is consistent with our result of no statistically significant evolution ($0.004 \pm 0.061\,\rm pc\, cm^{-3}\, yr^{-1}$ at 95\% confidence level).
Recently, using high-time-resolution CHIME data and temporally narrow bursts, \cite{Abbott26} report that FRB 20220912A shows a linear increase in DM of $1.4 \pm 0.6\ {\rm pc\ cm^{-3}\ yr^{-1}}$ (2.3$\sigma$), which is close to our result of $1.41 \pm 0.53\ {\rm pc\ cm^{-3}\ yr^{-1}}$ at the 95\% confidence level.

In the study of \citet{Curtin25}, six repeating FRBs overlap with our sample, namely FRB 20181119A, FRB 20190208A, FRB 20190303A, FRB 20190417A, FRB 20191106C, and FRB 20200929C.
Among these sources, FRB 20190208A, FRB 20190417A, and FRB 20191106C are consistent with our results in showing no significant DM evolution when their 1$\sigma$ uncertainties are converted to our 95\% confidence level for comparison.
For FRB 20190303A, one golden-sample source in our analysis, both obtains negative best-fit slopes, but with different uncertainties (both converted to the 95\% confidence level): $-0.04\pm0.039\, {\rm pc\ cm^{-3}\ yr^{-1}}$  
\citep[original is $-0.04\pm0.02\, {\rm pc\ cm^{-3}\ yr^{-1}}$ at 1$\sigma$;][]{Curtin25} versus $-0.23\pm 0.13\ {\rm pc\ cm^{-3}\ yr^{-1}}$ in our work.
A possible reason for this discrepancy is used the different data period: our analysis covers approximately seven years, compared to about four and a half years in that work.
For the six overlapping sources, \citet{Curtin25} report an average of 9.3 bursts per source, which is lower than our average of 24.3 bursts and also below our selection threshold of ten bursts.
In addition, we apply an EFAC factor to account for potentially underestimated measurement errors, which propagates into the fittings and leads to larger uncertainties.
The remaining overlapping sources are not identified significant DM variations and classified as golden samples in our analysis.

\begin{figure}[h]
\plotone{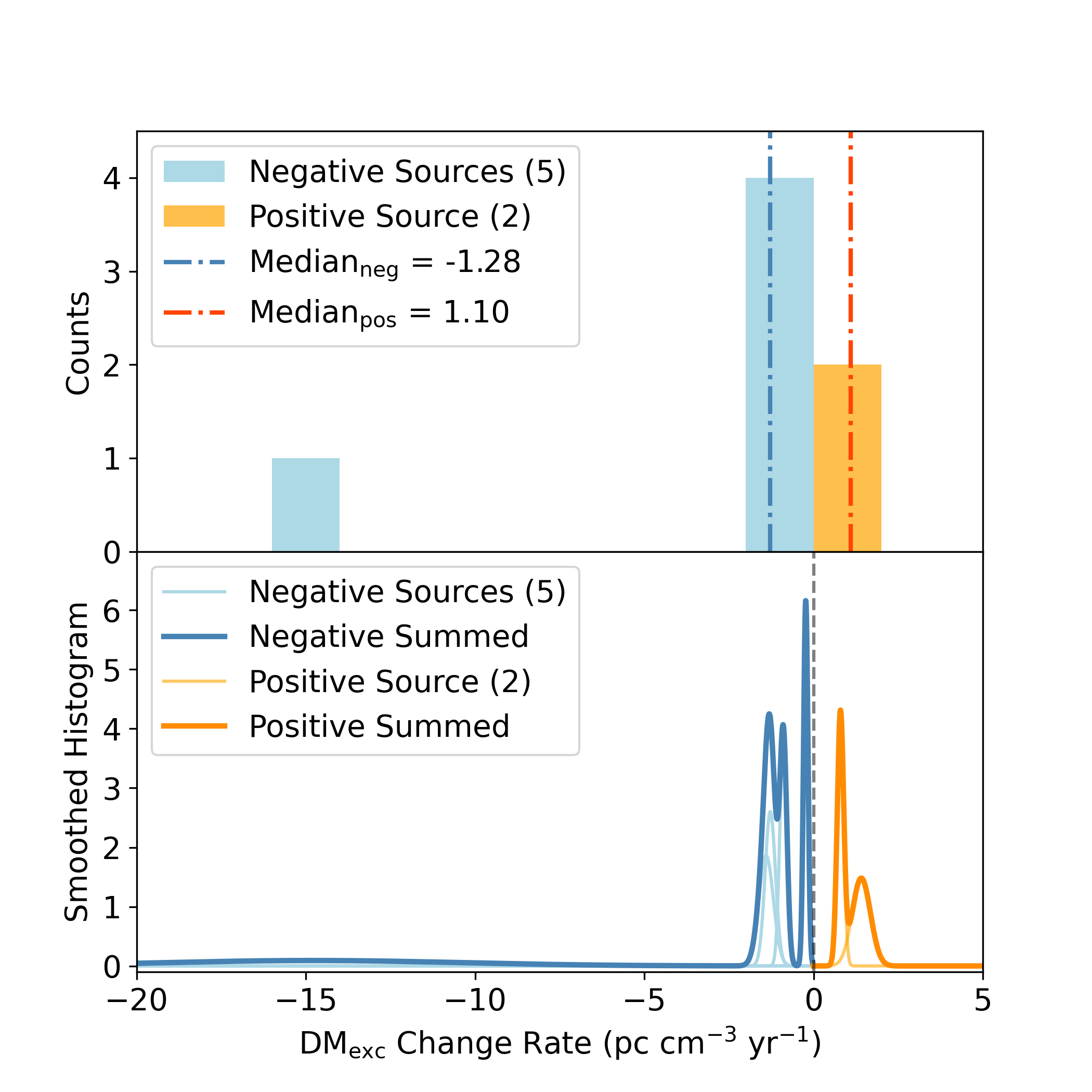}
\caption{Histogram of the annual ${\rm DM_{exc}}$ variation rates for the golden sample.
Upper panel: Blue and orange bars represent sources with decreasing and increasing DM, respectively. 
Blue and red dash-dotted lines mark the corresponding median DM variation rates.
Lower panel: Smoothed histograms that consider the uncertainties of individual DM variation rates. 
The light blue and light orange curves show the probability distributions of the DM variation rate for each source, while the dark blue and orange curves represent the summed distributions for the decreasing and increasing populations, respectively.}
\label{rate_hist}
\end{figure}

\subsection{Uncertainty in $\rm DM$ Trends}\label{3.3}

In this work, we use the DMs reported by CHIME (Section  \ref{2}), including both S/N-maximizing and structure-dependent methods\footnote{https://www.chime-frb.ca/analysis}.
\citet{Bilous2025} showed that in the latter method, absorption of the downward drift effect seen in repeaters gets absorbed into to DM, for low-S/N bursts.
In other words, downward drifting morphology may become degenerate with dispersive delays and thus bias DM measurements \citep{FengY26}.
In such a case, the DM trend reported above might potentially be caused, by a secular evolution in burst S/N, fully or in part.
To test this, we performed a Pearson correlation analysis between S/N and DM for the seven golden-sample repeaters.
Among these sources, only one (FRB 20181128A) exhibits a Pearson correlation coefficient larger than 0.2, with a value of 0.23.
The corresponding p-value is 0.384, which is well above the significance threshold of 0.05, indicating that the null hypothesis of zero correlation cannot be rejected.
So, the apparent weak correlation may arise from random fluctuations.
For the other golden sample sources, the Pearson correlation coefficients are all below 0.2, and all p-values are larger than 0.05.
Taken together, we find no statistical evidence for a correlation between S/N and DM in the golden sample.
This suggests that the observed DM variation trends are unlikely to be driven by S/N-dependent biases.

The remaining uncertainties then arise from instrumental effects at different epochs \citep{CHIME21a}. 
To mitigate such effects, we apply a $\chi^2$-based EFAC uncertainty inflation to the DM measurement errors.
Combined with a Bayesian analysis as a robustness test, we restrict our subsequent analysis to the golden samples.
Nevertheless, it is important to note that instrumental or selection effects may still exist. 

We note that in pulsar timing analyses an EQUAD parameter is 
sometimes used in addition to the EFAC, for handling error underestimations \citep{Iraci2024}.
This EQUAD parameter is a common uncertainty factor that is added in quadrature to all measurement uncertainties, until 
the reduced $\chi^2$ becomes unity.
However, our datasets are much smaller than pulsar timing ones, making the EQUAD parameter difficult to constrain and potentially drowning out any real DM evolution.
Thus, here we only apply an EFAC parameter to account for underestimated measurement errors, while noting that the EFAC-only approach will generally provide an optimistic estimate of the uncertainties on the fitted parameters.

Isolated points biased by potential instrumental effects may also influence the fitting.
For example, in FRB 20240316A shown in Figure \ref{frb_slope_grid_neg}, a single burst detected at an early epoch in 2024 (the leftmost data point) has a noticeable influence on the fitted DM rate.
There is, however, no clear justification for excluding this data point from the analysis. 
Its reported DM error is arguably relatively small ($\sim0.4\ \mathrm{pc\ cm^{-3}}$), but it is within the range of the errors reported for the other points, as visible in Figure \ref{frb_slope_grid_neg}. 
Its reported S/N is 7.6, which is comparable to other detected bursts. 
Except for its detection date, this point does not stand out from the other points.
Thus, we think down weighing this point either manually or through further change to the fitting are unjustified.  
However, it is worth noting that the single leftmost point of FRB 20240316A has the highest leverage on the fitted slope, and its DM rate uncertainty is almost certainly underestimated. 
This could represent a worst-case scenario for the EFAC-only approach.

Furthermore, from a logical standpoint and considering the sample as a whole, sources exhibiting increasing and decreasing DM trends are detected by the same instrument with large overlapping epochs.
If a significant instrumental bias were present, it would be expected to affect all sources in a similar manner, rather than producing systematically increasing DM trends in some repeaters but decreasing trends in the others.
This consideration provides additional support that the observed DM variation trends is unlikely to be solely driven by instrumental effects.

Based on the statistical analysis presented in this section, two main results emerge.
First, as discussed in Section \ref{2}, the observed variations in ${\rm DM_{exc}}$ are expected to be dominated by the local environment ${\rm DM_{local}}$.
The presence of both increasing and decreasing DM trends among repeating FRBs suggests diversity in the local environment evolution, or existence of multiple repeater populations.
Second, as illustrated by the histogram in Figure~\ref{rate_hist}, a larger fraction of repeaters exhibit an overall decreasing trend in ${\rm DM_{exc}}$.
This apparent population asymmetry indicates that decreasing DM trends are more common than increasing ones among CHIME repeating FRBs.

\section{Discussion} \label{4}

In this section, we perform a statistical test on apparent population asymmetry in DM variation trends.
We then explore possible physical interpretations based on supernova remnant (SNR) expansion and discuss potential methods for constraining local DM contributions.

\subsection{Are decreasing DM trends more common than increasing ones?}\label{4.1}
Based on the previous statistics, the decreasing DM population apparently outnumbers those with increasing DM within the golden sample.
We perform a binomial test under the null hypothesis of equal probabilities for both DM trends to examine whether decreasing ones are statistically more common in the repeater population. 
Using the full golden sample (5 decreasing and 2 increasing sources), we obtain a p-value of 0.23.
Considering that the DM rate fits for FRB 20240316A and FRB 20220912A may be affected by isolated data points, we exclude these two sources to get 4 decreasing and 1 increasing sources.
Then, the p-value is 0.19. 
These results indicate that, although sources with decreasing DM appear to outnumber those with increasing DM in the our golden sample, the conclusion that decreasing trends dominate the repeater population remains only marginally significant.

Given that several repeaters have been well monitored, and there is no evidence for selection bias in choosing sources (i.e., when two FRBs both become active, one is selected for observation while the other is not),
we further combine our golden sample with additional sources reported in the publications shown significant DM evolution. 
Specifically, we include FRB 20180301A \citep{Kumar23}, FRB 20121102A \citep{WangP25, Snelders25}, FRB 20190520B \citep{NiuCH26}, and FRB 20220529A \citep{Pandhi26}. 
The combined sample includes 9 decreasing and 2 increasing sources, and its binomial test gives a p-value of 0.033, indicating that decreasing DM sources are statistically more common than increasing ones in repeater population.

\begin{figure}[]
\plotone{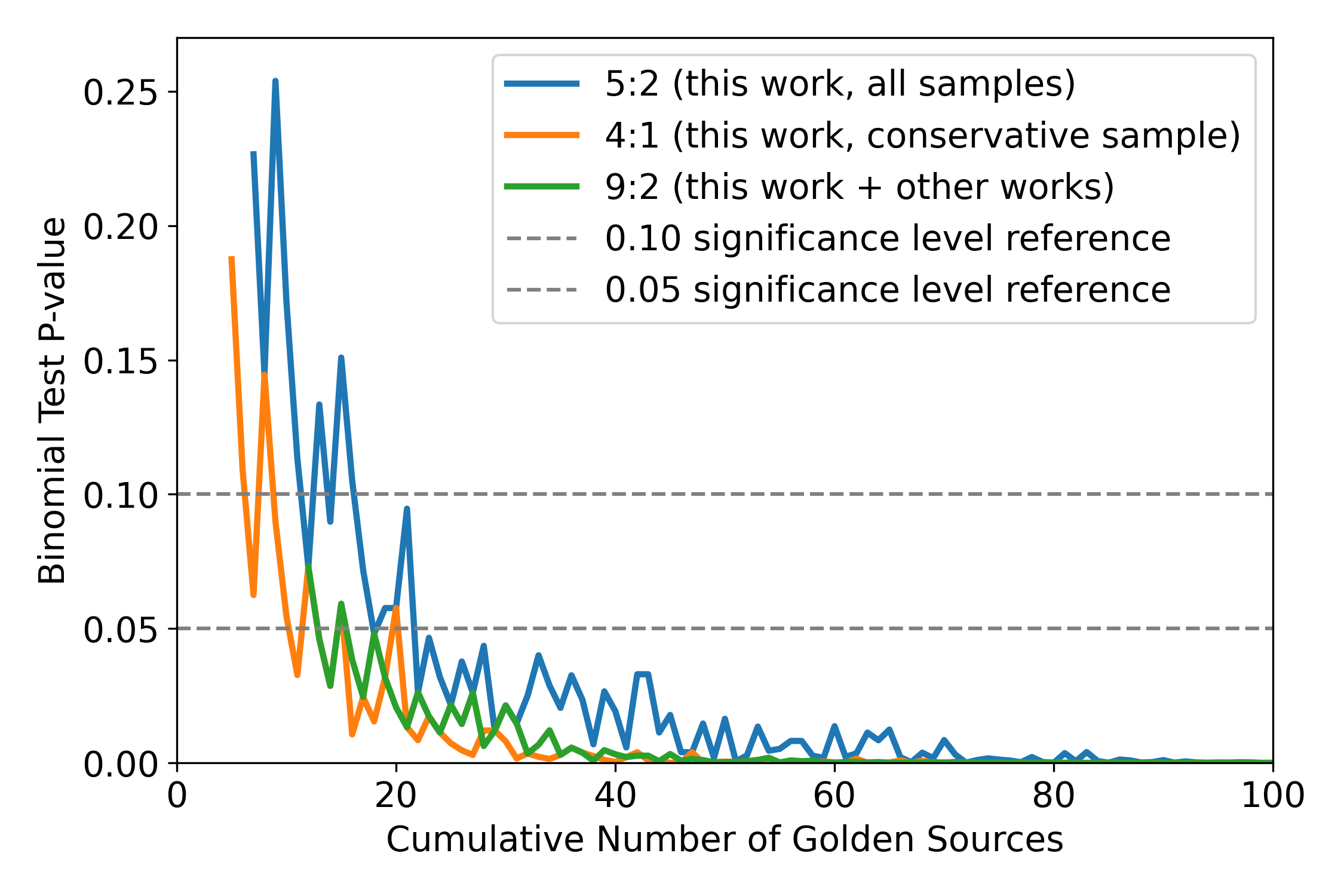}
\caption{
Variation of the binomial test p-value with increasing number of golden sources.
The blue, orange, and green curves represent different initial ratios of decreasing to increasing sources, respectively.
The 5:2 case uses all golden sample sources in this work. 
The 4:1 caes excludes two sources in this work that may be affected by isolated data points in the fit, that are FRB 20240316A and FRB 20220912A.
The 9:2 case corresponds to the combined sample of this work golden samples and other studies, including FRB 20180301A \citep{Kumar23}, FRB 20121102A \citep{WangP25, Snelders25}, FRB 20190520B \citep{NiuCH26}, and FRB 20220529A \citep{Pandhi26}.}
\label{bino_p}
\end{figure}

The differences in the p-values suggest that statistical tests based on small samples may be affected by limited data size.
We therefore to examine how the p-value evolves with increasing sample size based on the current observed number ratios.
With the initial numbers of increasing ($\rm I_{0}$) and decreasing ($\rm D_{0}$) sources, we use a Beta distribution to get the probability distribution as $\mathrm{Beta}(\rm 1+I_0,\,1+D_0)$ under an uniform prior.
This Beta distribution is fixed by the initial sample and is used to predict future cases.
We then randomly draw a probability from this Beta distribution to generate the numbers of increasing and decreasing sources in the future sample. 
These are combined with the initial sample, and a binomial test is performed to obtain the corresponding p-value.
For each future sample size, this process loops 1000 times, and the median of the p-values represents the result for that sample size.
We apply this procedure to three cases as shown in Figure \ref{bino_p}: (i) our full golden sample, with $\rm I_{0}=2$ and $\rm D_{0}=5$; (ii) a conservative subsample excluding sources potentially affected by isolated points, with $\rm I_{0}=1$ and $\rm D_{0}=4$; and (iii) a combined sample including additional sources reported by the literature, with $\rm I_{0}=2$ and $\rm D_{0}=9$.
In all three cases, the p-value decreases as the sample size increases. 
Even for the most conservative case, the p-value converges below 0.05 when the total sample size reaches about 30.

Taken together, based on the above discussions, the statistical test for the current combined sample supports the conclusion that decreasing DM trends are more common than increasing DM trends among repeating FRBs.
Further simulations based on the number ratios (5:2) given by the current full CHIME sample show that, when the number of repeaters with accurately measured DM rates reaches approximately 30, it becomes possible to determine whether decreasing DM trends outnumber increasing ones with statistical significance.


\subsection{Supernova remnant expansion as an explanation for decreasing $\rm DM$}\label{4.2}
Given the prevalence of a decreasing DM trend among repeating FRBs, a natural interpretation is a decline in the electron density of their local environments.
The SNR expansion model provides a promising explanation for this picture \citep{YangYP17, Piro18}.
Recent long-term observations of two active repeating FRBs (FRB20190520B and FRB20121102A) have reported significant decreases in local DM \citep{WangP25, NiuCH26}, suggesting that these sources reside in expanding ionized media like young SNRs.
The expansion of SNRs and their interactions with the surrounding ISM have been studied in detail.
Here, we build upon previous studies to further discuss the contributions of SNRs to the DM.

\subsubsection{A brief introduction of the SNR expansion model}
In general, the expansion of a SNR can be described by two evolutionary stages \citep{McKee95}.
At early times, the dynamics are dominated by the SN ejecta, and the remnant expands approximately at a constant velocity.
As the swept-up ISM mass becomes comparable to the ejecta mass, the expansion transitions to a decelerating phase.
As the SNR expands faster than the swept-up ISM, forward and reverse shocks are formed. 
This transition occurs at the Sedov–Taylor timescale ($t_{\rm ST}$), given by \citep{Sedov46, Taylor50}
\begin{equation}
    t_{\rm ST} = 210E_{51}^{-1/2}M_{1\odot}^{5/6}n_0^{-1/3}\, \rm years,
\label{tst}
\end{equation}
where $E_{51}$ is the explosion energy in units of $10^{51}$ erg, $M_{1\odot}$ is the ejecta mass in solar masses, and $n_0=1\,\rm cm^{-3}$ is the standard surrounding ISM number density \citep{Piro18}.

\begin{figure}[h]
\plotone{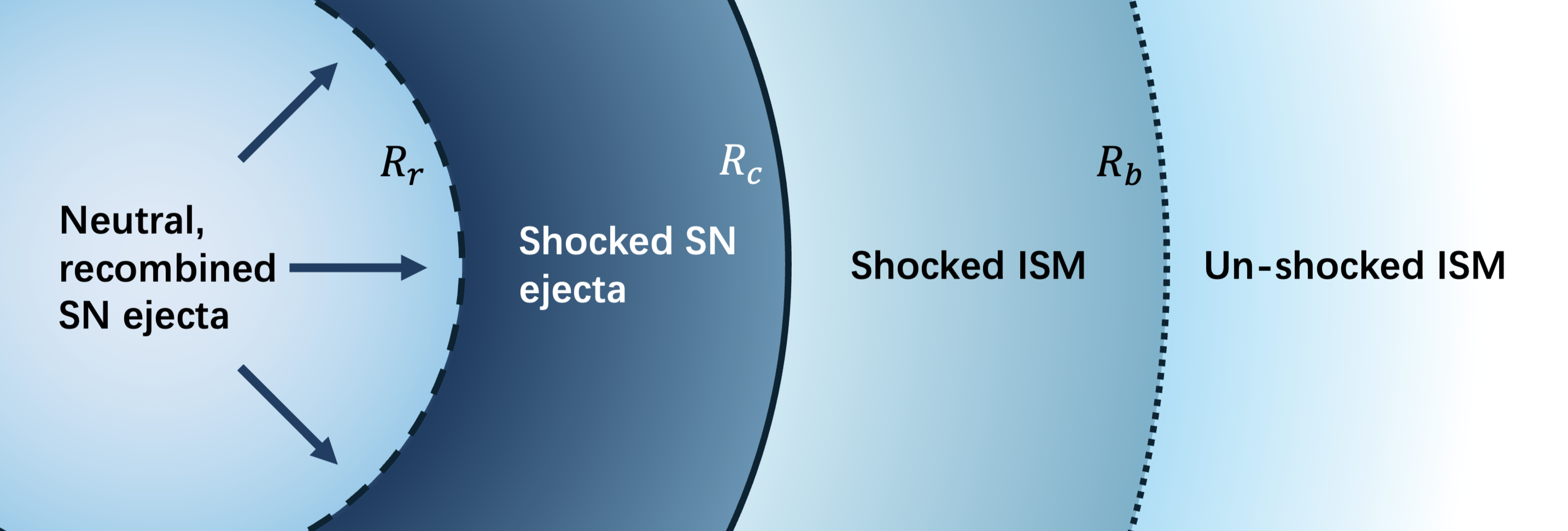}
\caption{Schematic illustration of the SNR expansion. 
$R_{\rm r}$, $R_{\rm c}$, and $R_{\rm b}$ represent the radii of the reverse shock, the contact discontinuity, and the shocked ISM, respectively, adapted from \cite{Piro18}.}
\label{SNR_sch}
\end{figure}
From a spatial perspective, the SNR structure consists of a shocked ejecta region bounded by the reverse shock ($R_r$) and the contact discontinuity ($R_c$), surrounded by the shocked and unshocked ISM, as illustrated in Figure \ref{SNR_sch}.
For young remnants ($t \lesssim t_{\rm ST}$), the electron density in the shocked ejecta is significantly higher than that in the surrounding regions, and thus dominates the contribution to the dispersion measure \citep{Piro18}.
The DM contribution from the SNR can therefore be approximated as
\begin{equation}
     {\rm DM_{SNR}} = \int^{R_{c}}_{R_{r}}n_{rc} {\rm d}l\approx(R_{c}-R_{r})n_{rc},
\label{dm_snr}
\end{equation}
where $n_{rc}$ is the electron number density between the reverse shock and the contact discontinuity.
In the $n_{rc}$ dominated region, the density is higher than the average density of the remnant, but the pressure is continuous, which can be written as \citep{Piro18}
\begin{equation}
     n_{ rc}=4n_0\frac{\mu}{\mu_e}\left(\frac{v_{b}}{v_{r}}\right)^2,
\label{ncr}
\end{equation}
where $\mu$ and $\mu_e$ are the mean molecular weight and the mean molecular weight per electron, respectively, and $v_b$ and $v_r$ are the velocities of the forward and reverse shocks.\citep{Piro18}.
By substituting Eq. \ref{ncr} into Eq. \ref{dm_snr} and using the expressions of $R_{c}$, $R_{r}$, $v_{b}$, and $v_{r}$ for $t \le t_{\rm ST}$ and $t > t_{\rm ST}$ (see Appendix \ref{B}), we can get evolution of ${\rm DM_{SNR}}$.
Because CHIME operates from 400 MHz to 800 MHz band, FRB signals become increasingly difficult to detect once the plasma frequency of the SNR media exceeds 400 MHz \citep{ShangLH17}. 
Therefore, we mark the time when the electron density evolves to the point where the plasma frequency decays to 400 MHz.

\subsubsection{$\rm DM_{SNR}$ estimation}
Based on the SNR expansion model described above, we estimate the evolutionary age of FRBs and its contribution to DM.
Due to the limited size of the golden sample, a population analysis is not yet feasible, so we focus on outlining a methodology that can be applied to larger samples in future.
The basic idea is to infer the evolutionary age of each FRB by comparing its observed DM decreasing rate with that predicted by the model.
The inferred age can then be used to estimate the corresponding DM contribution from the SNR.

In Figure \ref{evo_combined}, we demonstrate this procedure using the five golden-sample sources currently available.
The best-fit DM rate of each source is indicated by a horizontal orange dashed line, and the blue solid curve shows the model predicted DM rate.
The intersections between the horizontal lines and the model curve are marked by red stars, whose sizes reflect the burst rates of the sources.
The light-red shaded regions represent the uncertainties of the estimated ages, propagated from the measurement uncertainties of the DM rates.
Combining the inferred ages with the model predicted DM evolution, we estimate ${\rm DM_{SNR}}$ for each source at its corresponding evolutionary age in the lower panel of Figure \ref{evo_combined}.

\begin{figure}[ht]
\plotone{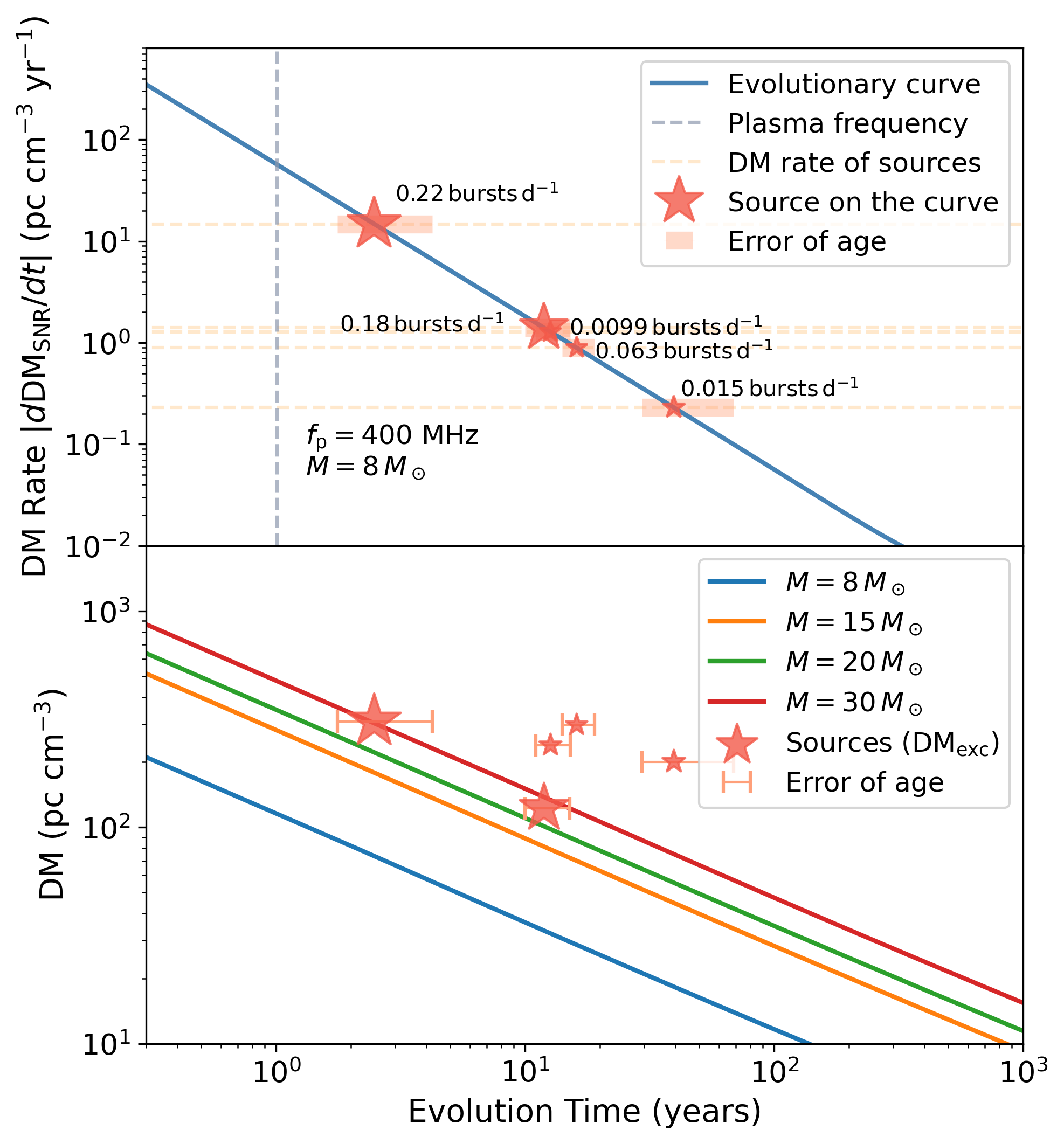}
\caption{Illustration of FRB estimated evolutionary ages and ${\rm DM_{SNR}}$ contributions within the SNR framework.  
Upper panel: Orange horizontal dashed lines show the $\rm DM_{exc}$ best-fit rate for each source.
The star size is reflecting burst rate.
The blue solid curve represents the SNR model assuming a progenitor mass of $8M_\odot$.
The light red shaded regions (vertical extent has no physical meaning) indicate the uncertainties in the inferred evolutionary ages for each source, propagated from the 95\% confidence of the measured DM variation rates in Section \ref{3}.
Lower panel: The solid curves in different colors represent SNR models assuming different progenitor masses.
The horizontal position of each red star indicates the inferred evolutionary age of the source, and the corresponding vertical position on the model curve gives the estimated ${\rm DM_{SNR}}$ contribution at that age.}
\label{evo_combined}
\end{figure}

\newpage

\subsection{Diversity in $\rm DM$ Trends}\label{4.3}

Our statistical analysis shows that in the golden sample a minority displays an increasing trend.
This indicates that not all sources can be explained by one picture or evolutionary stage \citep{LinHH24}. 
Our results further extend this view, implying that such diversity may also exist within the repeater population itself \citep{CuiXH25}, potentially arising from different birth processes or local environments \citep{ZhangB23}.

The binary accretion–triggered FRB model represents another competitive class of repeater scenarios, in which the FRB source is typically assumed to be a magnetar or neutron star in a binary system with a massive companion \citep{XuH22, WangFY22}.
Recently, \cite{LiY2026} reported rapid changes in the local plasma environment of a repeating FRB 20220529, and interpreted them as potentially arising from binary orbital modulation.
In interacting binary systems, material from the companion can be transferred via stellar winds or Roche-lobe overflow, forming an accretion flow \citep{Frank02}.
As the accretion process evolves, the column density of electrons along the line of sight changes. 
An increase in DM is expected to occur near the periastron, where the stellar wind density is higher or Roche-lobe overflow becomes significant.
For systems with long orbital periods, this can manifest as an overall increase trend in DM on year-timescales. 
In this case, DM variations are fundamentally modulated by the orbital phase, and a subsequent decrease may also occur.

Another possible explanation invokes specific SNR-related effects.
If the neutron star receives a kick velocity during an anisotropic core-collapse explosion \citep{Frail94, Janka17}, it may move away from the observer and pass through the far side of the SNR shell.
During this passage, the electron density along the line of sight can increase, potentially leading to a rise in the observed DM.
Given that the SNR shell can span several parsecs, such an effect may produce an increasing DM trend over much longer timescales.

\section{Conclusion} \label{5}
Among the 63 repeaters reported by CHIME, we examine the 19 most frequently detected repeating FRBs in order to conduct a statistical analysis regarding DM variations.
Of these, seven sources (the ``golden sample") meet our criteria for DM evolution significance.
Within this golden sample, we identify two types of DM evolution trends: an overall decrease (five sources with a median DM rate of $-1.28 \, \rm pc\, cm^{-3}\, yr^{-1}$) and an overall increase (two sources with a median DM rate of $1.10 \, \rm pc\, cm^{-3}\, yr^{-1}$).
These two evolutionary patterns may provide clues for the diversity among repeating FRBs or their local surroundings.

A decreasing DM trend is more common than an increasing one among the currently known repeating FRBs, with a binomial test p-value of 0.033. 
If future discoveries follow the currently observed increasing-to-decreasing number ratios, this p-value will naturally decrease further.
We conclude that the current observational evidence already provides an indication that DMs generally decrease in the population of repeating FRBs. 


We interpret the dominance of the decreasing DM trend within the framework of the young SNR expansion model.
By comparing the measured DM variation rates with model expectations, we derive illustrative estimates of the evolutionary ages of individual repeaters and their corresponding $\rm DM_{SNR}$ contributions.

\section*{Acknowledgments}
We thank Bryan M. Gaensler for valuable suggestions on DM errors and SNR models, Gopakumar Achamveedu for insightful discussions on Galactic radio pulsars, Jumei Yao for helpful discussions on the DM contributions from the Milky Way and its halo, and Lingqi Meng for useful discussions on pulsar timing method.
We are 
grateful to the anonymous referee for their constructive comments on this paper.
This work is supported by NSFC No. 12588202 and U2031117, 
National Key R\&D Program of China No. 2023YFE0110500, 
the International Partnership Program of Chinese Academy of Sciences No.114A11KYSB20210010, 
grant NSF PHY-2309135 to the Kavli Institute for Theoretical Physics (KITP), 
the Youth Innovation Promotion Association CAS (id. 2021055),
the CAS Youth Interdisciplinary Team No. QN2023061004L,
and CORTEX (NWA.1160.18.316) financed by the Dutch Research Council (NWO).
Erbil Gugercinoglu is supported by the Doctor Foundation of Qingdao Binhai University (No. BJZA2025025).


\appendix

\section{Characteristics of selected repeating FRBs}\label{A}
\setcounter{table}{0}
\renewcommand{\thetable}{A.\arabic{table}}

Based on CHIME observations, we selected 19 repeating FRBs with more than 10 detected bursts from 63 repeaters. 
Their basic and derived properties are summarized in Table \ref{frb-dmrate}.

\begin{table*}
\centering
\caption{Repeating FRBs analyzed in this work and their DM variation rates.}
\label{frb-dmrate}
\begin{tabular}{lccccccccc}
\hline
FRB & Burst & \multicolumn{1}{c}{RA} & \multicolumn{1}{c}{Dec} & \multicolumn{1}{c}{$l$} & \multicolumn{1}{c}{$b$} & \multicolumn{1}{c}{Mean $\rm DM_{exc}$} &
\multicolumn{1}{c}{WLS DM rate} & \multicolumn{1}{c}{Bayesian DM rate} & Golden \\
Name & Number & \multicolumn{1}{c}{(deg)} & \multicolumn{1}{c}{(deg)} & \multicolumn{1}{c}{(deg)} & \multicolumn{1}{c}{(deg)} & \multicolumn{1}{c}{(pc cm$^{-3}$)} &
\multicolumn{1}{c}{(pc cm$^{-3}$ yr$^{-1}$)} & \multicolumn{1}{c}{(pc cm$^{-3}$ yr$^{-1}$)} & Sample \\
\hline
FRB 20220912A & 445 & 209.05 & 48.70 & 157.05 & 61.64 & 201.4 &
$1.41 \pm 0.53$ & $1.41\pm 0.53$ & Yes \\

FRB 20180916B & 169 & 29.50 & 65.73 & 129.71 & 3.74 & 24.7 &
$0.004 \pm 0.061$ & $0.004^{+0.055}_{-0.053}$ & No \\

FRB 20240209A & 127 & 289.89 & 86.07 & 118.57 & 26.58 & 122.7 &
$-1.40 \pm 0.42$ & $-1.39\pm0.41$ & Yes \\

FRB 20190303A & 38 & 208.25 & 48.25 & 97.48 & 65.72 & 201.0 &
$-0.23 \pm 0.13$ & $-0.22\pm0.12$ & Yes \\

FRB 20201124A & 34 & 77.00 & 26.05 & 177.77 & $-8.54$ & 217.7 &
$7.50 \pm 8.84$ & $7.17^{+8.85}_{-8.88}$ & No \\

FRB 20240316A$^{a}$ & 34 & 354.58 & 32.38 & 105.47 & $-28.00$ & 308.6 &
$-14.78 \pm 8.62$ & $-15.16^{+8.18}_{-8.51}$ & Yes \\

FRB 20181119A & 33 & 190.50 & 65.13 & 124.54 & 51.97 & 340.0 &
$0.15 \pm 0.17$ & $0.15^{+0.16}_{-0.17}$ & No \\

FRB 20190117A & 27 & 331.75 & 17.38 & 76.35 & $-30.25$ & 357.0 &
$0.80 \pm 0.19$ & $0.80\pm0.19$ & Yes \\

FRB 20190417A & 23 & 294.75 & 59.40 & 91.46 & 17.41 & 1298.0 &
$-0.12 \pm 0.40$ & $-0.10^{+0.42}_{-0.41}$ & No \\

FRB 20180814A & 22 & 65.50 & 73.67 & 136.42 & 16.60 & 81.2 &
$1.51 \pm 2.28$ & $2.88\pm2.67$ & No \\

FRB 20191106C & 20 & 199.50 & 43.00 & 105.88 & 73.24 & 310.7 &
$0.34 \pm 0.45$ & $0.32^{+0.53}_{-0.54}$ & No \\

FRB 20210323C & 17 & 122.00 & 72.33 & 142.59 & 31.54 & 239.6 &
$-1.28 \pm 0.30$ & $-1.28^{+0.33}_{-0.32}$ & Yes \\

FRB 20181128A & 16 & 74.00 & 63.38 & 146.62 & 12.43 & 297.4 &
$-0.89 \pm 0.20$ & $-0.89^{+0.20}_{-0.21}$ & Yes \\

FRB 20190208A & 16 & 283.75 & 46.97 & 76.78 & 18.90 & 513.6 &
$-0.11 \pm 0.13$ & $-0.12\pm 0.14$ & No \\

FRB 20200223B & 16 & 8.25 & 28.82 & 118.07 & $-33.88$ & 165.0 &
$0.05 \pm 0.34$ & $0.05\pm0.35$ & No \\

FRB 20200929C & 16 & 17.00 & 18.47 & 128.42 & $-44.23$ & 385.0 &
$0.05 \pm 0.47$ & $0.06^{+0.49}_{-0.51}$ & No \\

FRB 20200120E & 14 & 149.25 & 68.82 & 142.26 & 41.15 & 55.7 &
$-0.03 \pm 0.06$ & $-0.02\pm0.17$ & No \\

FRB 20190212A & 13 & 276.00 & 81.43 & 113.31 & 27.82 & 253.1 &
$-0.05 \pm 0.08$ & $-0.05\pm0.18$ & No \\

FRB 20201130A & 12 & 64.25 & 7.93 & 185.33 & $-29.15$ & 220.1 &
$0.68 \pm 6.19$ & $0.42^{+6.93}_{-6.66}$ & No \\
\hline
\end{tabular}
\begin{flushleft}
Notes. Data are cited from https://www.chime-frb.ca/repeaters.
DM variation rates are measured using weighted least-squares (WLS) and Bayesian methods, with uncertainties corresponding to 95\% confidence level.
$^{a}$ The uncertainty of the fitted rate for this source is likely to be underestimated due to a high leverage isolated point and limitations of the EFAC-only approach.
\end{flushleft}
\end{table*}
\section{Expressions of SNR expansion}\label{B}
\setcounter{figure}{0}
\renewcommand{\thefigure}{B.\arabic{figure}}

In Section 4.1, we physically interpret the decreasing DM trend using the SNR expansion model \citep{YangYP17, Piro18}. We adopt the basic expressions for SNR evolution from \cite{Piro18}, assuming a surrounding ISM with constant density.
Figure \ref{SNR_sch} shows that during the evolution of a supernova remnant (SNR), it possesses distinct shells, each expanding at a different velocity.
The expansion velocities of the reverse shock ($v_{r}$) and the shocked ISM ($v_{b}$) before and after $t_{\rm ST}$ are,
\begin{equation}
    v_{r} = 
\begin{cases}
1.14v_{\rm ST}\left(\frac{t}{t_{\rm ST}}\right)^{3/2} \left[1+1.13\left(\frac{t}{t_{\rm ST}}\right)^{3/2} \right]^{-5/3}, & t\le t_{\rm ST}\\
\\
v_{\rm ST}\left[0.37 + 0.03\left(\frac{t}{t_{\rm ST}} \right) \right], & t> t_{\rm ST}, \\
\end{cases}
\label{vr}
\end{equation}
and,
\begin{equation}
    v_{b} = 
\begin{cases}
1.37v_{\rm ST}\left[1+0.60\left(\frac{t}{t_{\rm ST}}\right)^{3/2} \right]^{-5/3}, & t\le t_{\rm ST}\\
\\
0.63v_{\rm ST}\left[1.56\left(\frac{t}{t_{\rm ST}} \right)-0.56 \right]^{-3/5}, & t> t_{\rm ST}. \\
\end{cases}
\label{vb}
\end{equation}
Here, at the time $t_{\rm ST}$, the characteristic scale and velocity are given by,
\begin{equation}
    R_{\rm ST} = 2.2M_{1\odot}^{1/3}n_0^{-1/3}\ \rm pc,
\label{r_st}
\end{equation}
\begin{equation}
    v_{\rm ST} = 10^{4}E_{51}^{1/2}M_{1\odot}^{-1/2}\ \rm km \ s^{-1}.
\label{c_st}
\end{equation}
At this stage, the electron density in the shocked SN ejecta (Eq. \ref{ncr}), which contributes the main component of the electron column density, can be expressed as:
\begin{equation}
    n_{rc} = 
\begin{cases}
3.77n_0\left(\frac{\mu}{\mu_e}\right)\left(\frac{t}{t_{\rm ST}}\right)^{-3}\left( \frac{1+0.60(t/t_{\rm ST})^{3/2}}{1+1.13(t/t_{\rm ST})^{3/2}} \right)^{-10/3}, & t\le t_{\rm ST} \\
\\
4n_0\left(\frac{\mu}{\mu_e}\right)\left(\frac{0.63\left[1.56(t/t_{\rm ST})-0.56\right]^{-3/5}}{0.37+0.03(t/t_{\rm ST})}\right)^2, & t> t_{\rm ST} \\
\end{cases}
\label{ncr-2}
\end{equation}
Meanwhile, $R_{c}$ and $R_{r}$ are written as,
\begin{equation}
    R_{c} = 
\begin{cases}
1.24R_{\rm ST}\left(\frac{t}{t_{\rm ST}}\right)\left[ 1+0.60\left(\frac{t}{t_{\rm ST}}\right)^{3/2}\right]^{-2/3}, & t\le t_{\rm ST} \\
\\
0.91R_{\rm ST}\left[1.56\left(\frac{t}{t_{\rm ST}}\right)-0.56\right]^{2/5}, & t> t_{\rm ST}, \\
\end{cases}
\label{rc-2}
\end{equation}
and,
\begin{equation}
    R_{r} = 
\begin{cases}
1.24R_{\rm ST}\left(\frac{t}{t_{\rm ST}}\right)\left[ 1+1.13\left(\frac{t}{t_{\rm ST}}\right)^{3/2}\right]^{-2/3}, & t\le t_{\rm ST} \\
\\
R_{\rm ST}\left(\frac{t}{t_{\rm ST}}\right)\left[0.78-0.03\left(\frac{t}{t_{\rm ST}}\right)-0.37{\rm ln}\left(\frac{t}{t_{\rm ST}}\right)\right], & t> t_{\rm ST}. \\
\end{cases}
\label{rr-2}
\end{equation}
Then, by substituting the above expressions into Eq. \ref{dm_snr}, we can obtain the $\rm DM_{SNR}$ evolution with time, as
\begin{equation}
    {\rm DM_{SNR}} (T) = 
\begin{cases}
4.6748\, R_{\rm ST} n_0 \frac{\mu}{\mu_e} T^{-2}
\left( [1 + 0.60 T^{3/2}]^{-2/3} - [1 + 1.13 T^{3/2}]^{-2/3} \right)
\left( \frac{1+0.60 T^{3/2}}{1+1.13 T^{3/2}} \right)^{-10/3}, & T \le 1 \\
\\
\left[ 0.91 R_{\rm ST} [1.56 T - 0.56]^{2/5} -R_{\rm ST} T (0.78 - 0.03 T - 0.37 \ln T) \right]
4 n_0 \frac{\mu}{\mu_e} \left( \frac{0.63 [1.56 T - 0.56]^{-3/5}}{0.37 + 0.03 T} \right)^2, & T > 1,
\end{cases}
\label{dm-snr-2}
\end{equation}
where $T$ is defined as $T = t / t_{\rm ST}$.
The temporal evolution of $\rm DM_{SNR}$ and its first derivative (the $\rm DM_{SNR}$ variation rate) are shown in Figure \ref{evo_curve_dm_ddm}, where a standard ISM density of $n_0 = 1\ \rm cm^{-3}$ is assumed throughout.

\begin{figure*}
\plotone{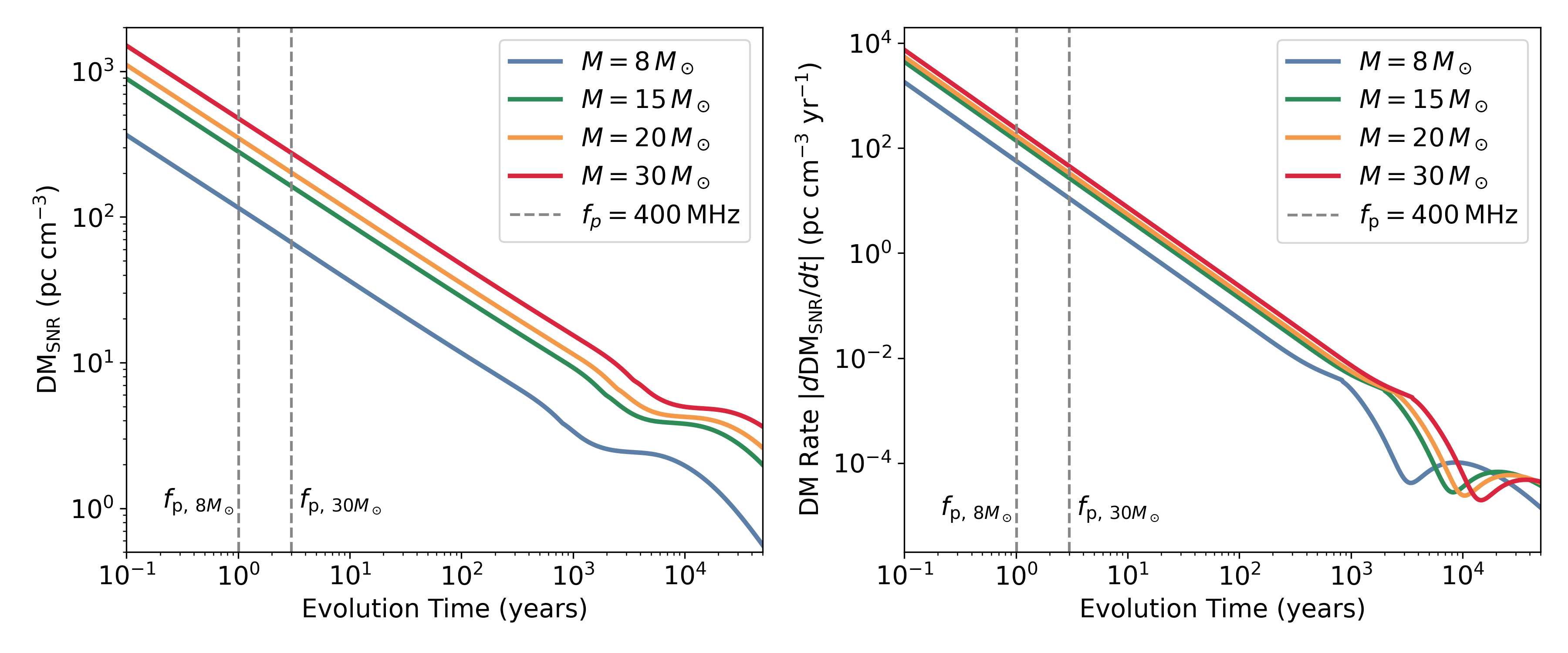}
\caption{Evolution of $\rm DM_{SNR}$ (left) and its variation rate (right) for different progenitor star masses.
Vertical dashed lines indicate the evolutionary stage corresponding to a plasma frequency of 400 MHz.}
\label{evo_curve_dm_ddm}
\end{figure*}

\bibliography{reference2025}{}

\begin{thebibliography}{}
\expandafter\ifx\csname natexlab\endcsname\relax\def\natexlab#1{#1}\fi
\providecommand{\url}[1]{\href{#1}{#1}}
\providecommand{\dodoi}[1]{doi:~\href{http://doi.org/#1}{\nolinkurl{#1}}}
\providecommand{\doeprint}[1]{\href{http://ascl.net/#1}{\nolinkurl{http://ascl.net/#1}}}
\providecommand{\doarXiv}[1]{\href{https://arxiv.org/abs/#1}{\nolinkurl{https://arxiv.org/abs/#1}}}

\bibitem[{{Abbott} {et~al.}(2026){Abbott}, {Pearlman}, {Kaspi}, {Pandhi}, {Brar}, {Cassity}, {Cook}, {Curtin}, {Fonseca}, {Gaensler}, {Good}, {Hessels}, {Khan}, {Leung}, {Main}, {Mckinven}, {Meyers}, {Nimmo}, {Ng}, {Pleunis}, {Scholz}, {Shah}, \& {Shin}}]{Abbott26}
{Abbott}, T.~C., {Pearlman}, A.~B., {Kaspi}, V.~M., {et~al.} 2026, arXiv e-prints, arXiv:2604.09098, \dodoi{10.48550/arXiv.2604.09098}

\bibitem[{Backer {et~al.}(2000)Backer, Wong, \& Valanju}]{Backer_2000}
Backer, D.~C., Wong, T., \& Valanju, J. 2000, The Astrophysical Journal, 543, 740, \dodoi{10.1086/317150}

\bibitem[{{Bailes}(2022)}]{Bailes22}
{Bailes}, M. 2022, Science, 378, abj3043, \dodoi{10.1126/science.abj3043}

\bibitem[{{Bilous} {et~al.}(2025){Bilous}, {van Leeuwen}, {Maan}, {Pastor-Marazuela}, {Oostrum}, {Rajwade}, \& {Wang}}]{Bilous2025}
{Bilous}, A.~V., {van Leeuwen}, J., {Maan}, Y., {et~al.} 2025, \aap, 696, A194, \dodoi{10.1051/0004-6361/202451413}

\bibitem[{{Bochenek} {et~al.}(2020){Bochenek}, {Ravi}, {Belov}, {Hallinan}, {Kocz}, {Kulkarni}, \& {McKenna}}]{Bochenek20}
{Bochenek}, C.~D., {Ravi}, V., {Belov}, K.~V., {et~al.} 2020, \nat, 587, 59, \dodoi{10.1038/s41586-020-2872-x}

\bibitem[{{Bruni} {et~al.}(2024){Bruni}, {Piro}, {Yang}, {Quai}, {Zhang}, {Palazzi}, {Nicastro}, {Feruglio}, {Tripodi}, {O'Connor}, {Gardini}, {Savaglio}, {Rossi}, {Nicuesa Guelbenzu}, \& {Paladino}}]{Bruni24}
{Bruni}, G., {Piro}, L., {Yang}, Y.-P., {et~al.} 2024, \nat, 632, 1014, \dodoi{10.1038/s41586-024-07782-6}

\bibitem[{{Bruni} {et~al.}(2025){Bruni}, {Piro}, {Yang}, {Palazzi}, {Nicastro}, {Rossi}, {Savaglio}, {Maiorano}, \& {Zhang}}]{Bruni25}
{Bruni}, G., {Piro}, L., {Yang}, Y.~P., {et~al.} 2025, \aap, 695, L12, \dodoi{10.1051/0004-6361/202453233}

\bibitem[{{Chatterjee} {et~al.}(2017){Chatterjee}, {Law}, {Wharton}, {Burke-Spolaor}, {Hessels}, {Bower}, {Cordes}, {Tendulkar}, {Bassa}, {Demorest}, {Butler}, {Seymour}, {Scholz}, {Abruzzo}, {Bogdanov}, {Kaspi}, {Keimpema}, {Lazio}, {Marcote}, {McLaughlin}, {Paragi}, {Ransom}, {Rupen}, {Spitler}, \& {van Langevelde}}]{Chatterjee17}
{Chatterjee}, S., {Law}, C.~J., {Wharton}, R.~S., {et~al.} 2017, \nat, 541, 58, \dodoi{10.1038/nature20797}

\bibitem[{{CHIME/FRB Collaboration} {et~al.}(2018){CHIME/FRB Collaboration}, {Amiri}, {Bandura}, {Berger}, {Bhardwaj}, {Boyce}, {Boyle}, {Brar}, {Burhanpurkar}, {Chawla}, {Chowdhury}, {Cliche}, {Cranmer}, {Cubranic}, {Deng}, {Denman}, {Dobbs}, {Fandino}, {Fonseca}, {Gaensler}, {Giri}, {Gilbert}, {Good}, {Guliani}, {Halpern}, {Hinshaw}, {H{\"o}fer}, {Josephy}, {Kaspi}, {Landecker}, {Lang}, {Liao}, {Masui}, {Mena-Parra}, {Naidu}, {Newburgh}, {Ng}, {Patel}, {Pen}, {Pinsonneault-Marotte}, {Pleunis}, {Rafiei Ravandi}, {Ransom}, {Renard}, {Scholz}, {Sigurdson}, {Siegel}, {Smith}, {Stairs}, {Tendulkar}, {Vanderlinde}, \& {Wiebe}}]{CHIME18}
{CHIME/FRB Collaboration}, {Amiri}, M., {Bandura}, K., {et~al.} 2018, \apj, 863, 48, \dodoi{10.3847/1538-4357/aad188}

\bibitem[{{CHIME/FRB Collaboration} {et~al.}(2020{\natexlab{a}}){CHIME/FRB Collaboration}, {Andersen}, {Bandura}, {Bhardwaj}, {Bij}, {Boyce}, {Boyle}, {Brar}, {Cassanelli}, {Chawla}, {Chen}, {Cliche}, {Cook}, {Cubranic}, {Curtin}, {Denman}, {Dobbs}, {Dong}, {Fandino}, {Fonseca}, {Gaensler}, {Giri}, {Good}, {Halpern}, {Hill}, {Hinshaw}, {H{\"o}fer}, {Josephy}, {Kania}, {Kaspi}, {Landecker}, {Leung}, {Li}, {Lin}, {Masui}, {McKinven}, {Mena-Parra}, {Merryfield}, {Meyers}, {Michilli}, {Milutinovic}, {Mirhosseini}, {M{\"u}nchmeyer}, {Naidu}, {Newburgh}, {Ng}, {Patel}, {Pen}, {Pinsonneault-Marotte}, {Pleunis}, {Quine}, {Rafiei-Ravandi}, {Rahman}, {Ransom}, {Renard}, {Sanghavi}, {Scholz}, {Shaw}, {Shin}, {Siegel}, {Singh}, {Smegal}, {Smith}, {Stairs}, {Tan}, {Tendulkar}, {Tretyakov}, {Vanderlinde}, {Wang}, {Wulf}, \& {Zwaniga}}]{CHIME20a}
{CHIME/FRB Collaboration}, {Andersen}, B.~C., {Bandura}, K.~M., {et~al.} 2020{\natexlab{a}}, \nat, 587, 54, \dodoi{10.1038/s41586-020-2863-y}

\bibitem[{{CHIME/FRB Collaboration} {et~al.}(2020{\natexlab{b}}){CHIME/FRB Collaboration}, {Amiri}, {Andersen}, {Bandura}, {Bhardwaj}, {Boyle}, {Brar}, {Chawla}, {Chen}, {Cliche}, {Cubranic}, {Deng}, {Denman}, {Dobbs}, {Dong}, {Fandino}, {Fonseca}, {Gaensler}, {Giri}, {Good}, {Halpern}, {Hessels}, {Hill}, {H{\"o}fer}, {Josephy}, {Kania}, {Karuppusamy}, {Kaspi}, {Keimpema}, {Kirsten}, {Landecker}, {Lang}, {Leung}, {Li}, {Lin}, {Marcote}, {Masui}, {McKinven}, {Mena-Parra}, {Merryfield}, {Michilli}, {Milutinovic}, {Mirhosseini}, {Naidu}, {Newburgh}, {Ng}, {Nimmo}, {Paragi}, {Patel}, {Pen}, {Pinsonneault-Marotte}, {Pleunis}, {Rafiei-Ravandi}, {Rahman}, {Ransom}, {Renard}, {Sanghavi}, {Scholz}, {Shaw}, {Shin}, {Siegel}, {Singh}, {Smegal}, {Smith}, {Stairs}, {Tendulkar}, {Tretyakov}, {Vanderlinde}, {Wang}, {Wang}, {Wulf}, {Yadav}, \& {Zwaniga}}]{CHIME20b}
{CHIME/FRB Collaboration}, {Amiri}, M., {Andersen}, B.~C., {et~al.} 2020{\natexlab{b}}, \nat, 582, 351, \dodoi{10.1038/s41586-020-2398-2}

\bibitem[{{CHIME/FRB Collaboration} {et~al.}(2021){CHIME/FRB Collaboration}, {Amiri}, {Andersen}, {Bandura}, {Berger}, {Bhardwaj}, {Boyce}, {Boyle}, {Brar}, {Breitman}, {Cassanelli}, {Chawla}, {Chen}, {Cliche}, {Cook}, {Cubranic}, {Curtin}, {Deng}, {Dobbs}, {Dong}, {Eadie}, {Fandino}, {Fonseca}, {Gaensler}, {Giri}, {Good}, {Halpern}, {Hill}, {Hinshaw}, {Josephy}, {Kaczmarek}, {Kader}, {Kania}, {Kaspi}, {Landecker}, {Lang}, {Leung}, {Li}, {Lin}, {Masui}, {McKinven}, {Mena-Parra}, {Merryfield}, {Meyers}, {Michilli}, {Milutinovic}, {Mirhosseini}, {M{\"u}nchmeyer}, {Naidu}, {Newburgh}, {Ng}, {Patel}, {Pen}, {Petroff}, {Pinsonneault-Marotte}, {Pleunis}, {Rafiei-Ravandi}, {Rahman}, {Ransom}, {Renard}, {Sanghavi}, {Scholz}, {Shaw}, {Shin}, {Siegel}, {Sikora}, {Singh}, {Smith}, {Stairs}, {Tan}, {Tendulkar}, {Vanderlinde}, {Wang}, {Wulf}, \& {Zwaniga}}]{CHIME21a}
---. 2021, \apjs, 257, 59, \dodoi{10.3847/1538-4365/ac33ab}

\bibitem[{{Connor} {et~al.}(2025){Connor}, {Ravi}, {Sharma}, {Ocker}, {Faber}, {Hallinan}, {Harnach}, {Hellbourg}, {Hobbs}, {Hodge}, {Hodges}, {Kosogorov}, {Lamb}, {Law}, {Rasmussen}, {Sherman}, {Somalwar}, {Weinreb}, {Woody}, \& {Konietzka}}]{Connor25}
{Connor}, L., {Ravi}, V., {Sharma}, K., {et~al.} 2025, Nature Astronomy, 9, 1226, \dodoi{10.1038/s41550-025-02566-y}

\bibitem[{{Cook} {et~al.}(2026){Cook}, {Shin}, {Pleunis}, {Fine}, {Jain}, {Bingham}, {Curtin}, {Eadie}, {Gaensler}, {Hessels}, {Leung}, {Main}, {Mulyk}, {Pandhi}, {Scholz}, {Siegel}, {Stenning}, {Abbott}, {Andersen}, {Bhardwaj}, {Cai}, {Chatterjee}, {Dong}, {Fonseca}, {Hewitt}, {Joseph}, {Kahinga}, {Lazda}, {Kaspi}, {Khan}, {Kharel}, {Mas-Ribas}, {Masui}, {McGregor}, {Michilli}, {Mckinven}, {Ng}, {Nimmo}, {Shivraj Patil}, {Pearlman}, {Sammons}, {Sand}, {Sedaei Oghani}, {Shah}, {Smith}, {Stairs}, \& {Zegmott}}]{Cook26}
{Cook}, A.~M., {Shin}, K., {Pleunis}, Z., {et~al.} 2026, arXiv e-prints, arXiv:2605.08410.
\newblock \doarXiv{2605.08410}

\bibitem[{{Cordes} \& {Chatterjee}(2019)}]{Cordes19}
{Cordes}, J.~M., \& {Chatterjee}, S. 2019, \araa, 57, 417, \dodoi{10.1146/annurev-astro-091918-104501}

\bibitem[{{Cui} {et~al.}(2025){Cui}, {James}, {Li}, \& {Zhang}}]{CuiXH25}
{Cui}, X.-H., {James}, C.~W., {Li}, D., \& {Zhang}, C.-M. 2025, \apj, 982, 158, \dodoi{10.3847/1538-4357/adbbcb}

\bibitem[{{Cui} {et~al.}(2021){Cui}, {Zhang}, {Wang}, {Zhang}, {Li}, {Peng}, {Zhu}, {Wang}, {Strom}, {Ye}, {Wang}, \& {Yang}}]{CuiXH21}
{Cui}, X.-H., {Zhang}, C.-M., {Wang}, S.-Q., {et~al.} 2021, \mnras, 500, 3275, \dodoi{10.1093/mnras/staa3351}

\bibitem[{{Cui} {et~al.}(2022){Cui}, {Zhang}, {Li}, {Zhang}, {Peng}, {Zhu}, {Strom}, {Wang}, {Wang}, {Wu}, {Wang}, \& {Yang}}]{CuiXH22}
{Cui}, X.-H., {Zhang}, C.-M., {Li}, D., {et~al.} 2022, \apss, 367, 66, \dodoi{10.1007/s10509-022-04093-y}

\bibitem[{Curtin {et~al.}(2025)Curtin, Sand, Pleunis, Jain, Kaspi, Michilli, Fonseca, Shin, Nimmo, Brar, Dong, Eadie, Gaensler, Herrera-Martin, Ibik, Joseph, Kaczmarek, Leung, Main, Masui, Mckinven, Mena-Parra, Ng, Pandhi, Pearlman, Rafiei-Ravandi, Sammons, Scholz, Smith, \& Stairs}]{Curtin25}
Curtin, A.~P., Sand, K.~R., Pleunis, Z., {et~al.} 2025, ApJ, 992, 206, \dodoi{10.3847/1538-4357/adf844}

\bibitem[{{Feng} {et~al.}(2026){Feng}, {Zhou}, {Zhang}, {Li}, {Fang}, {Xu}, {Xu}, \& {Xie}}]{FengY26}
{Feng}, Y., {Zhou}, D., {Zhang}, Y., {et~al.} 2026, \apj, 1004, 179, \dodoi{10.3847/1538-4357/ae731b}

\bibitem[{{Feng} {et~al.}(2022){Feng}, {Li}, {Yang}, {Zhang}, {Zhu}, {Zhang}, {Lu}, {Wang}, {Dai}, {Lynch}, {Yao}, {Jiang}, {Niu}, {Zhou}, {Xu}, {Miao}, {Niu}, {Meng}, {Qian}, {Tsai}, {Wang}, {Xue}, {Yue}, {Yuan}, {Zhang}, \& {Zhang}}]{FengY22}
{Feng}, Y., {Li}, D., {Yang}, Y.-P., {et~al.} 2022, Science, 375, 1266, \dodoi{10.1126/science.abl7759}

\bibitem[{{Fonseca} {et~al.}(2020){Fonseca}, {Andersen}, {Bhardwaj}, {Chawla}, {Good}, {Josephy}, {Kaspi}, {Masui}, {Mckinven}, {Michilli}, {Pleunis}, {Shin}, {Tendulkar}, {Bandura}, {Boyle}, {Brar}, {Cassanelli}, {Cubranic}, {Dobbs}, {Dong}, {Gaensler}, {Hinshaw}, {Landecker}, {Leung}, {Li}, {Lin}, {Mena-Parra}, {Merryfield}, {Naidu}, {Ng}, {Patel}, {Pen}, {Rafiei-Ravandi}, {Rahman}, {Ransom}, {Scholz}, {Smith}, {Stairs}, {Vanderlinde}, {Yadav}, \& {Zwaniga}}]{Fonseca20}
{Fonseca}, E., {Andersen}, B.~C., {Bhardwaj}, M., {et~al.} 2020, \apjl, 891, L6, \dodoi{10.3847/2041-8213/ab7208}

\bibitem[{{Frail} {et~al.}(1994){Frail}, {Goss}, \& {Whiteoak}}]{Frail94}
{Frail}, D.~A., {Goss}, W.~M., \& {Whiteoak}, J.~B.~Z. 1994, \apj, 437, 781, \dodoi{10.1086/175038}

\bibitem[{{Frank} {et~al.}(2002){Frank}, {King}, \& {Raine}}]{Frank02}
{Frank}, J., {King}, A., \& {Raine}, D.~J. 2002, {Accretion Power in Astrophysics: Third Edition}

\bibitem[{{FRB Collaboration} {et~al.}(2026){FRB Collaboration}, {Abbott}, {Andersen}, {Andrew}, {Bandura}, {Bhardwaj}, {Bhusare}, {Brar}, {Cassanelli}, {Chatterjee}, {Cliche}, {Cook}, {Curtin}, {Dobbs}, {Dong}, {Eadie}, {Eftekhari}, {Fonseca}, {Gaensler}, {Good}, {Halpern}, {Hessels}, {Ibik}, {Jain}, {Joseph}, {Kader}, {Kaspi}, {Khan}, {Kharel}, {Kumar}, {Landecker}, {Lang}, {Lanman}, {L'Argent}, {Lazda}, {Leung}, {Li}, {Lintott}, {Main}, {Masui}, {Mate}, {McGregor}, {Mckinven}, {Mena-Parra}, {Meyers}, {Michilli}, {Ng}, {Ng}, {Nimmo}, {Noble}, {Pandhi}, {Patil}, {Pearlman}, {Pen}, {Pleunis}, {Prochaska}, {Rafiei-Ravandi}, {Ransom}, {Renard}, {Sammons}, {Sand}, {Scholz}, {Shah}, {Shin}, {Siegel}, {Sirota}, {Smith}, {Stairs}, {Stenning}, {Tendulkar}, {Vanderlinde}, {Walmsley}, {Wang}, \& {Wulf}}]{CHIME26}
{FRB Collaboration}, {Abbott}, T., {Andersen}, B.~C., {et~al.} 2026, arXiv e-prints, arXiv:2601.09399, \dodoi{10.48550/arXiv.2601.09399}

\bibitem[{{Gardenier} {et~al.}(2021){Gardenier}, {Connor}, {van Leeuwen}, {Oostrum}, \& {Petroff}}]{Gardenier2021}
{Gardenier}, D.~W., {Connor}, L., {van Leeuwen}, J., {Oostrum}, L.~C., \& {Petroff}, E. 2021, \aap, 647, A30, \dodoi{10.1051/0004-6361/202039626}

\bibitem[{{Iraci} {et~al.}(2024){Iraci}, {Chalumeau}, {Tiburzi}, {Verbiest}, {Possenti}, {Shaifullah}, {Susarla}, {Krishnakumar}, {Lam}, {Cromartie}, {Kerr}, \& {Grie{\ss}meier}}]{Iraci2024}
{Iraci}, F., {Chalumeau}, A., {Tiburzi}, C., {et~al.} 2024, \aap, 692, A170, \dodoi{10.1051/0004-6361/202450740}

\bibitem[{{James} {et~al.}(2022){James}, {Prochaska}, {Macquart}, {North-Hickey}, {Bannister}, \& {Dunning}}]{James22}
{James}, C.~W., {Prochaska}, J.~X., {Macquart}, J.~P., {et~al.} 2022, \mnras, 509, 4775, \dodoi{10.1093/mnras/stab3051}

\bibitem[{{Janka}(2017)}]{Janka17}
{Janka}, H.-T. 2017, \apj, 837, 84, \dodoi{10.3847/1538-4357/aa618e}

\bibitem[{{Kumar} {et~al.}(2023){Kumar}, {Luo}, {Price}, {Shannon}, {Deller}, {Bhandari}, {Feng}, {Flynn}, {Jiang}, {Uttarkar}, {Wang}, \& {Zhang}}]{Kumar23}
{Kumar}, P., {Luo}, R., {Price}, D.~C., {et~al.} 2023, \mnras, 526, 3652, \dodoi{10.1093/mnras/stad2969}

\bibitem[{{Kuzmin} {et~al.}(2008){Kuzmin}, {Losovsky}, {Jordan}, \& {Smith}}]{Kuzmin08}
{Kuzmin}, A., {Losovsky}, B.~Y., {Jordan}, C.~A., \& {Smith}, F.~G. 2008, \aap, 483, 13, \dodoi{10.1051/0004-6361:20079211}

\bibitem[{{Lan} {et~al.}(2024){Lan}, {Zhao}, {Wei}, \& {Wang}}]{LanHT24}
{Lan}, H.-T., {Zhao}, Z.-Y., {Wei}, Y.-J., \& {Wang}, F.-Y. 2024, \apjl, 967, L44, \dodoi{10.3847/2041-8213/ad4ae8}

\bibitem[{{Li} {et~al.}(2018){Li}, {Wang}, {Qian}, {Krco}, {Jiang}, {Yue}, {Jin}, {Zhu}, {Pan}, {Nan}, \& {Dunning}}]{LiD18}
{Li}, D., {Wang}, P., {Qian}, L., {et~al.} 2018, IEEE Microwave Magazine, 19, 112, \dodoi{10.1109/MMM.2018.2802178}

\bibitem[{{Li} {et~al.}(2021){Li}, {Wang}, {Zhu}, {Zhang}, {Zhang}, {Duan}, {Zhang}, {Feng}, {Tang}, {Chatterjee}, {Cordes}, {Cruces}, {Dai}, {Gajjar}, {Hobbs}, {Jin}, {Kramer}, {Lorimer}, {Miao}, {Niu}, {Niu}, {Pan}, {Qian}, {Spitler}, {Werthimer}, {Zhang}, {Wang}, {Xie}, {Yue}, {Zhang}, {Zhi}, \& {Zhu}}]{LiD21}
{Li}, D., {Wang}, P., {Zhu}, W.~W., {et~al.} 2021, \nat, 598, 267, \dodoi{10.1038/s41586-021-03878-5}

\bibitem[{{Li} {et~al.}(2026){Li}, {Zhang}, {Yang}, {Tsai}, {Yang}, {Law}, {Anna-Thomas}, {Chen}, {Lee}, {Tang}, {Xiao}, {Xu}, {Yang}, {Chen}, {Feng}, {Li}, {Mckinven}, {Niu}, {Shin}, {Wang}, {Zhang}, {Zhang}, {Zhou}, {Zhu}, {Dai}, {Chang}, {Geng}, {Han}, {Hu}, {Li}, {Luo}, {Niu}, {Shi}, {Sun}, {Wu}, {Zhu}, {Jiang}, \& {Zhang}}]{LiY2026}
{Li}, Y., {Zhang}, S.~B., {Yang}, Y.~P., {et~al.} 2026, Science, 391, 280, \dodoi{10.1126/science.adq3225}

\bibitem[{{Lin} {et~al.}(2024){Lin}, {Scholz}, {Ng}, {Pen}, {Bhardwaj}, {Chawla}, {Curtin}, {Li}, {Newburgh}, {Reda}, {Sand}, {Tendulkar}, {Andersen}, {Bandura}, {Brar}, {Cassanelli}, {Cook}, {Dobbs}, {Dong}, {Eadie}, {Fonseca}, {Gaensler}, {Giri}, {Herrera-Martin}, {Hill}, {Kaczmarek}, {Kania}, {Kaspi}, {Khairy}, {Lanman}, {Leung}, {Masui}, {Mena-Parra}, {Meyers}, {Michilli}, {Milutinovic}, {Ordog}, {Pearlman}, {Pleunis}, {Rafiei-Ravandi}, {Rahman}, {Ransom}, {Sanghavi}, {Shin}, {Smith}, {Stairs}, {Stenning}, {Vanderlinde}, \& {Wulf}}]{LinHH24}
{Lin}, H.-H., {Scholz}, P., {Ng}, C., {et~al.} 2024, \apj, 975, 75, \dodoi{10.3847/1538-4357/ad779d}

\bibitem[{Lorimer \& Kramer(2012)}]{Lorimer12}
Lorimer, D., \& Kramer, M. 2012, Handbook of Pulsar Astronomy

\bibitem[{{Lorimer} {et~al.}(2007){Lorimer}, {Bailes}, {McLaughlin}, {Narkevic}, \& {Crawford}}]{Lorimer07}
{Lorimer}, D.~R., {Bailes}, M., {McLaughlin}, M.~A., {Narkevic}, D.~J., \& {Crawford}, F. 2007, Science, 318, 777, \dodoi{10.1126/science.1147532}

\bibitem[{{Luo} {et~al.}(2018){Luo}, {Lee}, {Lorimer}, \& {Zhang}}]{LuoR18}
{Luo}, R., {Lee}, K., {Lorimer}, D.~R., \& {Zhang}, B. 2018, \mnras, 481, 2320, \dodoi{10.1093/mnras/sty2364}

\bibitem[{{Macquart} \& {Ekers}(2018)}]{Macquart18}
{Macquart}, J.~P., \& {Ekers}, R. 2018, \mnras, 480, 4211, \dodoi{10.1093/mnras/sty2083}

\bibitem[{{Macquart} {et~al.}(2020){Macquart}, {Prochaska}, {McQuinn}, {Bannister}, {Bhandari}, {Day}, {Deller}, {Ekers}, {James}, {Marnoch}, {Os{\l}owski}, {Phillips}, {Ryder}, {Scott}, {Shannon}, \& {Tejos}}]{Macquart20}
{Macquart}, J.~P., {Prochaska}, J.~X., {McQuinn}, M., {et~al.} 2020, \nat, 581, 391, \dodoi{10.1038/s41586-020-2300-2}

\bibitem[{{McKee} \& {Truelove}(1995)}]{McKee95}
{McKee}, C.~F., \& {Truelove}, J.~K. 1995, \physrep, 256, 157, \dodoi{10.1016/0370-1573(94)00106-D}

\bibitem[{{McKee} {et~al.}(2018){McKee}, {Lyne}, {Stappers}, {Bassa}, \& {Jordan}}]{McKee18}
{McKee}, J.~W., {Lyne}, A.~G., {Stappers}, B.~W., {Bassa}, C.~G., \& {Jordan}, C.~A. 2018, \mnras, 479, 4216, \dodoi{10.1093/mnras/sty1727}

\bibitem[{{Michilli} {et~al.}(2018){Michilli}, {Seymour}, {Hessels}, {Spitler}, {Gajjar}, {Archibald}, {Bower}, {Chatterjee}, {Cordes}, {Gourdji}, {Heald}, {Kaspi}, {Law}, {Sobey}, {Adams}, {Bassa}, {Bogdanov}, {Brinkman}, {Demorest}, {Fernandez}, {Hellbourg}, {Lazio}, {Lynch}, {Maddox}, {Marcote}, {McLaughlin}, {Paragi}, {Ransom}, {Scholz}, {Siemion}, {Tendulkar}, {van Rooy}, {Wharton}, \& {Whitlow}}]{Michilli18}
{Michilli}, D., {Seymour}, A., {Hessels}, J.~W.~T., {et~al.} 2018, \nat, 553, 182, \dodoi{10.1038/nature25149}

\bibitem[{{Moroianu} {et~al.}(2025){Moroianu}, {Bhandari}, {Drout}, {Hessels}, {Hewitt}, {Kirsten}, {Marcote}, {Pleunis}, {Snelders}, {Sridhar}, {Bach}, {Bempong-Manful}, {Bezrukovs}, {Blaauw}, {Bray}, {Buttaccio}, {Chatterjee}, {Corongiu}, {Feiler}, {Gaensler}, {Gawro{\'n}ski}, {Giroletti}, {Ibik}, {Karuppusamy}, {Lazda}, {Leung}, {Lindqvist}, {Masui}, {Michilli}, {Nimmo}, {Ould-Boukattine}, {Pandhi}, {Paragi}, {Pearlman}, {Puchalska}, {Scholz}, {Shin}, {Sluman}, {Trudu}, {Williams-Baldwin}, \& {Yang}}]{Moroianu25}
{Moroianu}, A.~M., {Bhandari}, S., {Drout}, M.~R., {et~al.} 2025, arXiv e-prints, arXiv:2509.05174, \dodoi{10.48550/arXiv.2509.05174}

\bibitem[{{Nan} {et~al.}(2011){Nan}, {Li}, {Jin}, {Wang}, {Zhu}, {Zhu}, {Zhang}, {Yue}, \& {Qian}}]{NanRD11}
{Nan}, R., {Li}, D., {Jin}, C., {et~al.} 2011, International Journal of Modern Physics D, 20, 989, \dodoi{10.1142/S0218271811019335}

\bibitem[{{Niu} {et~al.}(2022){Niu}, {Aggarwal}, {Li}, {Zhang}, {Chatterjee}, {Tsai}, {Yu}, {Law}, {Burke-Spolaor}, {Cordes}, {Zhang}, {Ocker}, {Yao}, {Wang}, {Feng}, {Niino}, {Bochenek}, {Cruces}, {Connor}, {Jiang}, {Dai}, {Luo}, {Li}, {Miao}, {Niu}, {Anna-Thomas}, {Sydnor}, {Stern}, {Wang}, {Yuan}, {Yue}, {Zhou}, {Yan}, {Zhu}, \& {Zhang}}]{NiuCH22}
{Niu}, C.~H., {Aggarwal}, K., {Li}, D., {et~al.} 2022, \nat, 606, 873, \dodoi{10.1038/s41586-022-04755-5}

\bibitem[{{Niu} {et~al.}(2026){Niu}, {Li}, {Yang}, {Zhu}, {Zhang}, {Zhang}, {Du}, {Yao}, {Zheng}, {Wang}, {Feng}, {Zhang}, {Zhu}, {Yu}, {Jiang}, {Dai}, {Tsai}, {Chen}, {Hou}, {Niu}, {Wang}, {Miao}, {Li}, \& {Zhang}}]{NiuCH26}
{Niu}, C.-H., {Li}, D., {Yang}, Y.-P., {et~al.} 2026, Science Bulletin, 71, 76, \dodoi{10.1016/j.scib.2025.11.023}

\bibitem[{{Oostrum} {et~al.}(2020){Oostrum}, {Maan}, {van Leeuwen}, {Connor}, {Petroff}, {Attema}, {Bast}, {Gardenier}, {Hargreaves}, {Kooistra}, {van der Schuur}, {Sclocco}, {Smits}, {Straal}, {ter Veen}, {Vohl}, {Adams}, {Adebahr}, {de Blok}, {van den Brink}, {van Cappellen}, {Coolen}, {Damstra}, {van Diepen}, {Frank}, {Hess}, {van der Hulst}, {Hut}, {Ivashina}, {Loose}, {Lucero}, {Mika}, {Morganti}, {Moss}, {Mulder}, {Norden}, {Oosterloo}, {Orr{\'u}}, {de Reijer}, {Ruiter}, {Vermaas}, {Wijnholds}, \& {Ziemke}}]{Oostrum2020}
{Oostrum}, L.~C., {Maan}, Y., {van Leeuwen}, J., {et~al.} 2020, \aap, 635, A61, \dodoi{10.1051/0004-6361/201937422}

\bibitem[{{Pandhi} {et~al.}(2026){Pandhi}, {Nimmo}, {Andrew}, {Brar}, {Chatterjee}, {Cook}, {Curtin}, {Gaensler}, {Gawronski}, {Hessels}, {Kaspi}, {Khan}, {Kirsten}, {Lazda}, {Leung}, {Main}, {Masui}, {Mckinven}, {Michilli}, {Ng}, {Ould-Boukattine}, {Pearlman}, {Pleunis}, {Pollak}, {Pradeep E.~T.}, {Puchalska}, {Sammons}, {Scholz}, {Shah}, {Shin}, {Siegel}, \& {Smith}}]{Pandhi26}
{Pandhi}, A., {Nimmo}, K., {Andrew}, S., {et~al.} 2026, \apjl, 1000, L53, \dodoi{10.3847/2041-8213/ae52f8}

\bibitem[{{Petroff} {et~al.}(2022){Petroff}, {Hessels}, \& {Lorimer}}]{Petroff22}
{Petroff}, E., {Hessels}, J.~W.~T., \& {Lorimer}, D.~R. 2022, \aapr, 30, 2, \dodoi{10.1007/s00159-022-00139-w}

\bibitem[{{Petroff} {et~al.}(2013){Petroff}, {Keith}, {Johnston}, {van Straten}, \& {Shannon}}]{Petroff13}
{Petroff}, E., {Keith}, M.~J., {Johnston}, S., {van Straten}, W., \& {Shannon}, R.~M. 2013, \mnras, 435, 1610, \dodoi{10.1093/mnras/stt1401}

\bibitem[{{Piro} \& {Gaensler}(2018)}]{Piro18}
{Piro}, A.~L., \& {Gaensler}, B.~M. 2018, \apj, 861, 150, \dodoi{10.3847/1538-4357/aac9bc}

\bibitem[{{Rajwade} {et~al.}(2020){Rajwade}, {Mickaliger}, {Stappers}, {Morello}, {Agarwal}, {Bassa}, {Breton}, {Caleb}, {Karastergiou}, {Keane}, \& {Lorimer}}]{Rajwade20}
{Rajwade}, K.~M., {Mickaliger}, M.~B., {Stappers}, B.~W., {et~al.} 2020, \mnras, 495, 3551, \dodoi{10.1093/mnras/staa1237}

\bibitem[{{Ravi} {et~al.}(2022){Ravi}, {Law}, {Li}, {Aggarwal}, {Bhardwaj}, {Burke-Spolaor}, {Connor}, {Lazio}, {Simard}, {Somalwar}, \& {Tendulkar}}]{Ravi22}
{Ravi}, V., {Law}, C.~J., {Li}, D., {et~al.} 2022, \mnras, 513, 982, \dodoi{10.1093/mnras/stac465}

\bibitem[{{Sedov}(1946)}]{Sedov46}
{Sedov}, L.~I. 1946, Journal of Applied Mathematics and Mechanics, 10, 241

\bibitem[{{Shang} {et~al.}(2017){Shang}, {Zhang}, {Li}, {Wang}, {Wang}, {Wang}, {Pan}, {Yang}, \& {Zhi}}]{ShangLH17}
{Shang}, L.-H., {Zhang}, C.-M., {Li}, D., {et~al.} 2017, \apj, 849, 87, \dodoi{10.3847/1538-4357/aa932c}

\bibitem[{{Snelders} {et~al.}(2025){Snelders}, {Hessels}, {Huang}, {Sridhar}, {Marcote}, {Moroianu}, {Ould-Boukattine}, {Kirsten}, {Bhandari}, {Hewitt}, {Pelliciari}, {Rhodes}, {Anna-Thomas}, {Bach}, {Bempong-Manful}, {Bezrukovs}, {Bray}, {Buttaccio}, {Cognard}, {Corongiu}, {Feiler}, {Gawro{\'n}ski}, {Giroletti}, {Guillemot}, {Karuppusamy}, {Lindqvist}, {Nimmo}, {Possenti}, {Puchalska}, \& {Williams-Baldwin}}]{Snelders25}
{Snelders}, M.~P., {Hessels}, J.~W.~T., {Huang}, J., {et~al.} 2025, arXiv e-prints, arXiv:2510.11352, \dodoi{10.48550/arXiv.2510.11352}

\bibitem[{{Straal} {et~al.}(2020){Straal}, {Connor}, \& {van Leeuwen}}]{Straal20}
{Straal}, S.~M., {Connor}, L., \& {van Leeuwen}, J. 2020, \aap, 634, A105, \dodoi{10.1051/0004-6361/201833376}

\bibitem[{{Taylor}(1950)}]{Taylor50}
{Taylor}, G. 1950, Proceedings of the Royal Society of London Series A, 201, 159, \dodoi{10.1098/rspa.1950.0049}

\bibitem[{{Uttarkar} {et~al.}(2026){Uttarkar}, {Shannon}, {Gourdji}, {Deller}, {Dial}, {Glowacki}, {Bera}, {Gordon}, {Ryder}, {Tejos}, {Bhandari}, \& {Wang}}]{Uttarkar26}
{Uttarkar}, P.~A., {Shannon}, R.~M., {Gourdji}, K., {et~al.} 2026, \mnras, 545, staf1997, \dodoi{10.1093/mnras/staf1997}

\bibitem[{{Wang} {et~al.}(2022){Wang}, {Zhang}, {Dai}, \& {Cheng}}]{WangFY22}
{Wang}, F.~Y., {Zhang}, G.~Q., {Dai}, Z.~G., \& {Cheng}, K.~S. 2022, Nature Communications, 13, 4382, \dodoi{10.1038/s41467-022-31923-y}

\bibitem[{{Wang} {et~al.}(2025){Wang}, {Zhang}, {Yang}, {Zhou}, {Zhang}, {Feng}, {Zhao}, {Fang}, {Li}, {Zhu}, {Zhang}, {Wang}, {Huang}, {Luo}, {Han}, {Lee}, {Tsai}, {Dai}, {Gao}, {Zheng}, {Cao}, {Chen}, {Gugercinoglu}, {Jiang}, {Jing}, {Li}, {Li}, {Lu}, {Luo}, {Lyu}, {Miao}, {Niu}, {Niu}, {Qu}, {Wang}, {Wang}, {Wang}, {Wang}, {Wu}, {Wu}, {Weng}, {Xiao}, {Xu}, {Yao}, {Zhang}, {Zhao}, {Liu}, {Zhang}, {Zhou}, {Zhang}, \& {Zhu}}]{WangP25}
{Wang}, P., {Zhang}, J.~S., {Yang}, Y.~P., {et~al.} 2025, arXiv e-prints, arXiv:2507.15790, \dodoi{10.48550/arXiv.2507.15790}

\bibitem[{{Wang} \& {van Leeuwen}(2024)}]{Wang2024}
{Wang}, Y., \& {van Leeuwen}, J. 2024, \aap, 690, A377, \dodoi{10.1051/0004-6361/202450673}

\bibitem[{{Wang} {et~al.}(2023){Wang}, {Kurban}, {Zhou}, {Yu}, \& {Wang}}]{WangYB23}
{Wang}, Y.-B., {Kurban}, A., {Zhou}, X., {Yu}, Y.-W., \& {Wang}, N. 2023, \mnras, 524, 569, \dodoi{10.1093/mnras/stad1922}

\bibitem[{{Xu} {et~al.}(2022){Xu}, {Niu}, {Chen}, {Lee}, {Zhu}, {Dong}, {Zhang}, {Jiang}, {Wang}, {Xu}, {Zhang}, {Fu}, {Filippenko}, {Peng}, {Zhou}, {Zhang}, {Wang}, {Feng}, {Li}, {Brink}, {Li}, {Lu}, {Yang}, {Caballero}, {Cai}, {Chen}, {Dai}, {Djorgovski}, {Esamdin}, {Gan}, {Guhathakurta}, {Han}, {Hao}, {Huang}, {Jiang}, {Li}, {Li}, {Li}, {Li}, {Li}, {Liu}, {Luo}, {Men}, {Niu}, {Peng}, {Qian}, {Song}, {Stern}, {Stockton}, {Sun}, {Wang}, {Wang}, {Wang}, {Wang}, {Wu}, {Xiao}, {Xiong}, {Xu}, {Xu}, {Yang}, {Yang}, {Yao}, {Yi}, {Yue}, {Yu}, {Yu}, {Yuan}, {Zhang}, {Zhang}, {Zhang}, {Zhao}, {Zheng}, {Zhu}, \& {Zou}}]{XuH22}
{Xu}, H., {Niu}, J.~R., {Chen}, P., {et~al.} 2022, \nat, 609, 685, \dodoi{10.1038/s41586-022-05071-8}

\bibitem[{{Yang} \& {Zhang}(2017)}]{YangYP17}
{Yang}, Y.-P., \& {Zhang}, B. 2017, \apj, 847, 22, \dodoi{10.3847/1538-4357/aa8721}

\bibitem[{{Yao} {et~al.}(2017){Yao}, {Manchester}, \& {Wang}}]{YaoJM16}
{Yao}, J.~M., {Manchester}, R.~N., \& {Wang}, N. 2017, \apj, 835, 29, \dodoi{10.3847/1538-4357/835/1/29}

\bibitem[{{Zhang}(2023)}]{ZhangB23}
{Zhang}, B. 2023, Reviews of Modern Physics, 95, 035005, \dodoi{10.1103/RevModPhys.95.035005}

\bibitem[{{Zhang} {et~al.}(2026){Zhang}, {Niu}, {Zhu}, {Li}, {Wang}, {Wang}, {Feng}, {Li}, {Niu}, {Wang}, {Yu}, {Zhang}, \& {Zheng}}]{ZhangJL2026}
{Zhang}, J.-h., {Niu}, C.-H., {Zhu}, Y.-h., {et~al.} 2026, \apj, 996, 4, \dodoi{10.3847/1538-4357/ae20f6}

\bibitem[{{Zhang} {et~al.}(2025){Zhang}, {Wang}, {Wang}, {Wu}, {Li}, {Zhu}, {Zhang}, {Gao}, {Lee}, {Han}, {Tsai}, {Wang}, {Huang}, {Zou}, {Zhou}, {Lu}, {Xie}, {Fang}, {Cao}, {Miao}, {Zhu}, {Chen}, {Cheng}, {Ke}, {Zhang}, {Zhang}, {Cao}, {Tian}, {Wu}, {Zhang}, {Niu}, {Zhou}, {Xu}, {Wang}, {Chen}, {Chen}, {Cui}, {Feng}, {G{\"u}gercino{\u{g}}lu}, {Huang}, {Li}, {Li}, {Li}, {Lin}, {Liu}, {Luo}, {Luo}, {Niu}, {Qu}, {Qu}, {Menberu Tedila}, {Wang}, {Wang}, {Wang}, {Wang}, {Weng}, {Wu}, {Xu}, {Yang}, {Yang}, {Yew}, {Yu}, {Zhang}, \& {Zhao}}]{ZhangJS25}
{Zhang}, J.-S., {Wang}, T.-C., {Wang}, P., {et~al.} 2025, arXiv e-prints, arXiv:2507.14707, \dodoi{10.48550/arXiv.2507.14707}

\bibitem[{{Zhang} {et~al.}(2022){Zhang}, {Wang}, {Feng}, {Zhang}, {Li}, {Tsai}, {Niu}, {Luo}, {Yao}, {Zhu}, {Han}, {Lee}, {Zhou}, {Niu}, {Jiang}, {Wang}, {Zhang}, {Xu}, {Wang}, \& {Xu}}]{ZhangYK22}
{Zhang}, Y.-K., {Wang}, P., {Feng}, Y., {et~al.} 2022, Research in Astronomy and Astrophysics, 22, 124002, \dodoi{10.1088/1674-4527/ac98f7}

\bibitem[{{Zhang} {et~al.}(2023){Zhang}, {Li}, {Zhang}, {Cao}, {Feng}, {Wang}, {Qu}, {Niu}, {Zhu}, {Han}, {Jiang}, {Lee}, {Li}, {Luo}, {Niu}, {Tsai}, {Wang}, {Wang}, {Wu}, {Xu}, {Yang}, {Zhang}, {Zhou}, \& {Zhu}}]{ZhangYK23}
{Zhang}, Y.-K., {Li}, D., {Zhang}, B., {et~al.} 2023, \apj, 955, 142, \dodoi{10.3847/1538-4357/aced0b}

\end{thebibliography}
\bibliographystyle{aasjournal}

\end{document}